\documentclass{emulateapj}

\slugcomment{Submitted to ApJ, 3 January 2007, accepted 22 March 2007}
\shorttitle{Steady-state evolution of debris disks around A stars}
\shortauthors{Wyatt et al.}

\begin{document}

\title{Steady-state evolution of debris disks around A stars}

\author{M. C. Wyatt}
\affil{Institute of Astronomy, University of Cambridge,
  Madingley Road, Cambridge CB3 0HA, UK}
\email{wyatt@ast.cam.ac.uk}

\author{R. Smith}
\affil{Institute for Astronomy, Royal Observatory,
  Blackford Hill, Edinburgh EH9 3HJ, UK}

\author{K. Y. L. Su and G. H. Rieke}
\affil{Steward Observatory, University of Arizona, 933 N Cherry Ave., Tucson,
  AZ 85721, USA}

\author{J. S. Greaves}
\affil{Scottish Universities Physics Alliance, University of St. Andrews,
  Physics \& Astronomy, North Haugh, St Andrews KY16 9SS, UK}

\author{C. A. Beichman\altaffilmark{1} and G. Bryden}
\affil{Jet Propulsion Laboratory, 4800 Oak Grove Drive, Pasadena, CA 91109, USA}

\altaffiltext{1}{Michelson Science Center, California Institute of Technology, M/S 100-22,
  Pasadena, CA 91125, USA}

\begin{abstract}
In this paper a simple analytical model for the steady-state evolution of debris
disks due to collisions is confronted with Spitzer observations of dust
around main sequence A stars.
All stars are assumed to have planetesimal belts with a distribution of
initial masses and radii.
In the model disk mass is constant until the largest planetesimals reach
collisional equilibrium whereupon the mass falls off $\propto t_{age}^{-1}$.
Using parameters that are reasonable within the context of planet formation
models and observations of proto-planetary disks, the detection statistics and
trends seen at both 24 and 70 $\mu$m can be fitted well by the model.
While there is no need to invoke stochastic evolution or delayed
stirring to explain the detection statistics of dust around A stars,
the model is also consistent with a moderate rate of stochastic events.
Potentially anomalous systems are identified by their high ratio of observed dust
luminosity to the maximum permissible in the model given their radii and
ages, $f/f_{\rm{max}}$;
these are HD3003, HD38678, HD115892, and HD172555.
It is not clear if their planetesimals have unusual properties (e.g., high strength
or low eccentricity), or if their dust is a transient phenomenon.
There are also well-studied examples from the literature where transient phenomena
are favored (e.g., Vega, HD69830).
However, the overall success of our model, which assumes planetesimals in all belts
have the same strength, eccentricity and maximum size, suggests there is a large
degree of uniformity in the outcome of planet formation.
The distribution of the radii of the planetesimal belts, once corrected for
detection bias, is found to follow $N(r) \propto r^{-0.8 \pm 0.3}$
in the range 3-120 AU.
Since the inner edge of a belt is often attributed to an unseen
planet, this provides a unique constraint on the planetary systems of
A stars.
It is also shown that the effect of P-R drag on the inner edge of A star disks
may need to be considered for those close to the Spitzer detection threshold,
such as HD2262, HD19356, HD106591, and HD115892.
Predictions are made for the upcoming SCUBA-2 survey, including that at least
17 of the 100 A stars should be detectable above 2 mJy at 850 $\mu$m, illustrating
how this model can be readily applied to the interpretation of future surveys.
\end{abstract}

\maketitle

\section{Introduction}
\label{s:intro}
The asteroid and Kuiper belts lie in regions of the solar system that are stable
over long timescales to the gravitational perturbations of the planets (Lecar et al. 2001).
Dust produced in collisions between planetesimals in these belts inhabits
a broader distribution, but one that is strongly influenced by the same gravitational
perturbations (Dermott et al. 1994; Liou \& Zook 1999; Moro-Mart\'{i}n \& Malhotra 
2003).
These belts are thought to have started off with more than 200 times their present mass
(Stern 1996; Bottke et al. 2005) and it is speculated that the belts were depleted in the
same event that caused the Late Heavy Bombardment (LHB).
The LHB was a period of $\sim 100$ Myr duration that occurred
700 Myr after the solar system formed when the terrestrial planets were bombarded
with a large influx of comets and asteroids, possibly due to restructuring
of the planetary system in a dynamical instability (Gomes et al. 2005).
The evolution of these planetesimal belts since this event is thought to have been
relatively slow, with mass falling off due to collisional processing, but with large
spikes in the dust content of the inner solar system occurring when two large asteroids
collided (Nesvorn\'{y} et al. 2003; Farley et al. 2006).
Knowledge of the planetesimal and dust content of the solar system and how it
evolves thus provides a rich source of information about both the planetesimals
themselves and the planetary system in which they reside.

It is important to determine the evolution of the planetesimal belts around
other stars, since this may be equally illuminating as to the nature and evolution
of their planetary systems.
Surveys have been undertaken to determine this evolution, by measuring the
infrared flux from dust produced in the collisional destruction of those
planetesimals (Aumann et al. 1984).
The observations are used to derive basic properties of the dust belts, such their infrared 
luminosity $L_{\rm{ir}}$ (quoted as fractional luminosity, $f=L_{\rm{ir}}/L_\star$), and
the distance of the dust from the star (which is either the same as, or strongly dependent
on, the planetesimal belt radius $r$), and these are then compared with stellar properties,
such as age and spectral type.
Most studies have focussed on the evolution of dust luminosity, showing that
fractional luminosity decreases with age;
e.g., Spangler et al. (2001) found the mean dust luminosity of disks of similar age falls
off $\propto t^{-1.8}$, while Greaves \& Wyatt (2003) inferred from detection statistics
that the mass falls off $\propto t^{-0.5}$, Najita \& Williams (2005) found the
mass of the detected disks falls off $\propto t^{-1}$,
and the upper limit in 24 $\mu$m of the A star disks was found to fall off $\propto t^{-1}$
(Rieke et al. 2005), while disk mass from far-IR statistics was inferred to be
relatively age independent in the range 30-1000 Myr (Rhee et al. 2007).
These results are thus in general agreement with the results of theoretical models, 
which show that a planetesimal belt evolving in quasi-steady state would lose mass due 
to collisional
grinding down giving a disk mass (and dust luminosity) that falls off $\propto t^{-1}$
(Dominik \& Decin 2003; Wyatt et al. 2007).

However, Decin et al. (2003) noted that the maximum luminosity of debris disks
remains constant at $f \approx 10^{-3}$ even for the oldest stars, and this
was explained by Dominik \& Decin (2003) as a consequence of delayed stirring, in which
dust is not produced in the planetesimal belt until Pluto-sized objects form and
ignite a collisional cascade.
Since planet formation models predict that such massive bodies take longer to form
further from the star (Kenyon \& Bromley 2004), that interpretation makes the prediction
that the radius of the belts should increase with stellar age, and more recent surveys
have also considered the evolution of radius.
While in sub-mm surveys there is as yet no evidence of any evolution of radius with
age (Najita \& Williams 2005; Williams \& Andrews 2006), the latest far-IR results for
A stars do appear to show some evidence that radius is increasing with age, 
based on the fact that excess emission is detected at much later ages at 70 $\mu$m
than at 24 $\mu$m (Su et al. 2006).

It has also been suggested that much of the dust we see in some systems is produced
episodically in collisions between large planetesimals, much in the same way dust in
the zodiacal cloud increases when collisions occur in the asteroid belt, and that
such collisions contribute to the large spread in the infrared luminosities seen in disks
around A stars of similar age (Rieke et al. 2005).
This scenario is supported by the discovery of a population of dust grains around Vega in
the process of removal by radiation pressure which indicates that this system cannot have
remained in steady-state for the full 350 Myr age of the star (Su et al. 2005).
More recently it has also been shown that the few sun-like stars with hot dust at a few AU
(Gaidos 1999; Beichman et al. 2005; Song et al. 2005; Wyatt et al. 2005)
must be transient, since they are too bright for their age to be planetesimal belts
that have evolved due to steady-state collisional processing;
such systems may be undergoing periods analogous to the LHB in the solar system
(Wyatt et al. 2007).
A stochastic element to disk evolution would introduce additional free parameters
in models and would complicate interpretation of the observed statistics if it
plays a major role.

It has thus not yet been possible to arrive at a consistent picture for debris disk
evolution.
The fact that surveys are undertaken for different spectral types, at different
wavelengths, with different sensitivities to detections, has also added confusion
to the interpretation of the statistics.
In this paper we revisit a simple analytical model
for the steady-state collisional evolution of planetesimal belts which was
originally explored in Dominik \& Decin (2003) and recast with slightly modified
assumptions in Wyatt et al. (2007).
This model, including how it is applied in this paper to populations of
debris disks, is given in \S \ref{s:model}.
In \S \ref{s:rieke} we build a model for the population of disks
around A stars which is constrained using the 24 $\mu$m statistics presented in 
Rieke et al. (2005).
The properties of the model population are then compared in \S \ref{s:rsst05pred}
with those of A star disks detected in surveys at 70 $\mu$m (Su et al. 2006) and
the model parameters fine-tuned.
These results indicate that, with the exception of a few outlier systems, the
available statistics can be fitted by steady-state collisional evolution without
having to invoke a major role for stochasticity or delayed stirring.
Finally the model population is used to make predictions for future
debris disk surveys in the sub-mm (\S \ref{s:pred}).
The conclusions are given in \S \ref{s:conc}.

\section{Debris disk population model}
\label{s:model}
This section describes the simple analytical model for the
steady-state collisional evolution of planetesimal belts
that was derived in Wyatt et al. (2007) (\S \ref{ss:stst}),
and illustrates how that model is applied in this paper to
populations of debris disks (\S \ref{ss:pop}).

\subsection{Steady-state collisional evolution}
\label{ss:stst}
The planetesimal belt is assumed to be in collisional
equilibrium with a size distribution defined by:
\begin{equation}
  n(D) = K D^{2-3q}, \label{eq:nd}
\end{equation}
where $q$ is 11/6 for an infinite
collisional cascade (Dohnanyi 1969), and $D$ is the diameter of the planetesimals
in km.
That distribution is assumed to hold from the largest planetesimal
in the disk, of diameter $D_{\rm{c}}$ (in km), down to the size below which particles
are blown out by radiation pressure as soon as they are created,
$D_{\rm{bl}}$ (in $\mu$m).
If $q$ is in the range 5/3 to 2 then most of the
mass is in the largest planetesimals while the cross-sectional area is
dominated by the smallest particles such that for spherical particles of density $\rho$
in kg m$^{-3}$:
\begin{eqnarray}
  \sigma_{\rm{tot}} & = & 3.5 \times 10^{-17} K(3q-5)^{-1} (10^{-9}D_{\rm{bl}})^{5-3q}, \label{eq:stot} \\
  M_{\rm{tot}}      & = & 2.5 \times 10^{-9} (3q-5)(6-3q)^{-1} \rho \sigma_{\rm{tot}}
                         D_{\rm{bl}}(10^9 D_{\rm{c}}/D_{\rm{bl}})^{6-3q},
    \label{eq:mtot2}
\end{eqnarray}
where $\sigma_{\rm{tot}}$ is in AU$^2$ and $M_{\rm{tot}}$ is in $M_\oplus$ and the
rest of the units are as described above and are used consistently throughout the
paper.

The planetesimal belt is assumed to be at a radius, $r$, and have a width $dr$,
both in units of AU.
Assuming that the grains act like black bodies, and so absorb all the radiation
they intercept, the fractional luminosity, $f=L_{\rm{ir}}/L_\star$, of the
dust emission is given by
\begin{equation}
  f = \sigma_{\rm{tot}}/(4\pi r^2). \label{eq:f}
\end{equation}
In other words, in this model $\sigma_{\rm{tot}}$, $M_{\rm{tot}}$, and $f$ are
all proportional to each other and just one is needed to define the scaling
factor $K$ in equation (\ref{eq:nd}).

Assuming the particles act like black bodies also allows us to write down some
other simple relations:
\begin{equation}
  D_{\rm{bl}} = 0.8(L_\star/M_\star)(2700/\rho), \label{eq:dbl}
\end{equation}
where $L_\star$ and $M_\star$ are in solar units.
Also the emission from the disk at a wavelength $\lambda$ is given by:
\begin{equation}
  F_{\nu\rm{disk}} = 2.35 \times 10^{-11} B_\nu(\lambda,T)\sigma_{\rm{tot}}d^{-2}, 
  \label{eq:fnu}
\end{equation}
where d is the distance to the star in pc, giving $F_\nu$ in Jy if the Planck 
function $B_\nu$ is in Jy sr$^{-1}$.
The dust temperature can be worked out from:
\begin{equation}
  T = 278.3 L_\star^{0.25} r^{-0.5} \label{eq:tbb}
\end{equation}
in K.
The emission from the star at the same wavelength is given by:
\begin{equation}
  F_{\nu \star} = 1.77 B_\nu(\lambda,T_\star)L_\star T_\star^{-4}d^{-2},
  \label{eq:fnustar}
\end{equation}
where $T_\star$ is in K.

In a collisional cascade the amount of material within a given size range
$D$ to $D+dD$ decreases as these planetesimals are destroyed in collisions
with other members of the cascade, but is replaced at the same rate by
fragments created from the collisional destruction of larger objects. 
The long-timescale evolution is thus determined by the removal of mass from the top end of
the cascade.
In this model the scaling factor $K$ (and so the total mass and fractional luminosity)
decreases as the number of planetesimals of size $D_{\rm{c}}$ decreases.
The loss rate of such planetesimals is determined by their collisional lifetime,
which was derived in Wyatt et al. (2007) to be:
\begin{equation}
  t_{\rm{c}} =
        \left( \frac{3.8 \rho r^{3.5} (dr/r) D_{\rm{c}}}{M_\star^{0.5}M_{\rm{tot}}} \right) 
        \left( \frac{(12q-20)[1+1.25(e/I)^2]^{-0.5}}{(18-9q)G(q,X_{\rm{c}})} \right)
        \label{eq:tcmtot}
\end{equation}
in Myr, where $e$ and $I$ are the mean eccentricities and inclinations of the planetesimals'
orbits, and the factor $G(q,X_{\rm{c}})$ is given by
\begin{eqnarray}
  G(q,X_{\rm{c}}) & = & [(X_{\rm{c}}^{5-3q}-1)+ (6q-10)(3q-4)^{-1}(X_{\rm{c}}^{4-3q}-1) 
                   \nonumber \\
           &   & + (3q-5)(3q-3)^{-1}(X_{\rm{c}}^{3-3q}-1)], \label{eq:qgxc}
\end{eqnarray}
and $X_{\rm{c}}=D_{\rm{cc}}/D_{\rm{c}}$, where $D_{\rm{cc}}$ is the smallest planetesimal 
that has enough energy to catastrophically destroy a planetesimal of size $D_{\rm{c}}$.
The factor $X_{\rm{c}}$ can be worked out from the dispersal threshold, $Q_{\rm{D}}^\star$, defined
as the specific incident energy required to catastrophically destroy a particle such 
that (Wyatt \& Dent 2002):
\begin{equation}
  X_{\rm{c}} = 1.3 \times 10^{-3} \left[ \frac{Q_{\rm{D}}^\star r M_\star^{-1}}{1.25e^2+I^2} 
  \right]^{1/3}, 
\label{eq:xc}
\end{equation}
where $Q_{\rm{D}}^\star$ is in J kg$^{-1}$.

Assuming that collisions are the only process affecting the evolution of the disk
mass $M_{\rm{tot}}(t)$, then that mass (or equivalently for $K$, 
$\sigma_{\rm{tot}}$, or $f$) can be worked out by solving
$dM_{\rm{tot}}/dt = -M_{\rm{tot}}/t_{\rm{c}}$ to give:
\begin{equation}
  M_{\rm{tot}}(t) = M_{\rm{tot}}(0)/[1+t/t_{\rm{c}}(0)], \label{eq:mtott}
\end{equation}
where $M_{\rm{tot}}(0)$ is the initial disk mass and $t_{\rm{c}}(0)$ is the collisional 
lifetime at that initial epoch;
this solution is valid as long as mass is the only time variable property of the disk.
This results in a disk mass that is constant at $M_{\rm{tot}}(0)$ for
$t \ll t_{\rm{c}}(0)$, but that falls off $\propto t^{-1}$ for $t \gg t_{\rm{c}}(0)$.

As noted in Wyatt et al. (2007), one property of this evolution is that,
since the expression for $t_{\rm{c}}(0)$ includes a dependence on $M_{\rm{tot}}(0)$,
the disk mass at late times is independent of initial disk mass.
This is because more massive disks process their mass faster.
This means that for any given age, $t_{\rm{age}}$, there is a maximum disk mass
$M_{\rm{max}}$ (and also infrared luminosity, $f_{\rm{max}}$) that can remain due to 
collisional processing:
\begin{eqnarray}
  M_{\rm{max}} & = &
            \left( \frac{3.8 \times 10^{-6} \rho r^{3.5} (dr/r) D_{\rm{c}}}
                        {M_\star^{0.5} t_{\rm{age}}} \right) \times \nonumber \\
               &   & 
            \left( \frac{(12q-20)[1+1.25(e/I)^2]^{-0.5}}{(18-9q)G(q,X_{\rm{c}})} \right), 
            \label{eq:mmax1} \\
  f_{\rm{max}} & = &
            \left( \frac{10^{-6}r^{1.5}(dr/r)}{4\pi M_\star^{0.5} t_{\rm{age}}} \right)
            \times \nonumber \\ &   &
            \left( \frac{2[1+1.25(e/I)^2]^{-0.5}}{G(q,X_{\rm{c}})} \right)
            \left( \frac{D_{\rm{bl}}}{D_{\rm{c}}} \right)^{5-3q}, \label{eq:fmax1}
\end{eqnarray}
where again this mass is in units of $M_\oplus$.

\subsection{Application to populations of debris disks}
\label{ss:pop}
To apply this evolution to populations of debris disks, the
following assumptions were made.
All stars have a planetesimal belt that is undergoing a collisional cascade
(although this is not necessarily detectable).
The initial conditions of these belts are as described in \S \ref{ss:stst} using the
parameters $M_{\rm{tot}}(0)$, $r$, $dr$, $\rho$, $D_{\rm{c}}$, $D_{\rm{bl}}$ (which is 
defined by $L_\star$ and $M_\star$) and $q$.
The subsequent evolution is determined by $Q_{\rm{D}}^\star$, $e$ and $I$
(\S \ref{ss:stst}).
To simplify the problem we set $I=e$, $\rho=2700$ kg m$^{-3}$, $dr=r/2$
and $q=11/6$ for all disks and did not consider the effects of changing these parameters.
The parameters $Q_{\rm{D}}^\star$, $e$ and $D_{\rm{c}}$ were assumed to be the same for
all disks, and their values were constrained by a fit to the observed properties
of the known disks;
the consequence of choosing different values is discussed.
The remaining parameters, $M_{\rm{tot}}(0)$, $r$, $L_\star$ and $M_\star$ were chosen
from distributions for a population of a large number of disks (10,000 in this case).

For each disk the stellar spectral type was chosen at random from the
appropriate range (i.e., B8-A9 for comparison with
the sample of Rieke et al. 2005), thus defining $L_\star$, $M_\star$
and so $D_{\rm{bl}}$.
The distribution of initial disk masses, $M_{\rm{tot}}(0)$, was based on
the results of Andrews \& Williams (2005) who did a sub-mm study of proto-planetary
disks in the Taurus-Auriga star forming region, i.e., for $\sim 1M_\odot$ stars,
from which they derived that this population has a log-normal distribution of dust
masses that is centred on $M_{\rm{mid}} = 3.3M_\oplus$ and has a $1\sigma$ width of 1.14 dex.
We used the same distribution but allowed $M_{\rm{mid}}$ to be different from
that found in Taurus-Auriga, and this was a free parameter.
For the distribution of disk radii, a power law distribution
$N(r) \propto r^{\gamma}$ in the range 3-120 AU was adopted, where $N(r)dr$
is the number of disks with radii in the range $r$ to $r+dr$.
The range 3-120 AU was chosen from the range of radii inferred for the
sample of 46 A star disks with radius estimates (Table \ref{tab:2470sample}),
and the exponent $\gamma$ was determined from a fit to the observations.
The mass evolution of each disk was then completely defined using equations
(\ref{eq:tcmtot}) and (\ref{eq:mtott}), and its properties at the current
epoch were determined by assigning an age, $t_{\rm{age}}$, randomly from the
appropriate range (i.e., 0-800 Myr for A stars).
The stars were also assigned a random distance in the range 0-45 pc assuming an
isotropic distribution (i.e., $N(d) \propto d^2$);
given that there are $\sim 100$ A stars within 45 pc, the model
is overpopulated by a factor of around 100 compared with the real population
(although note that distance only becomes important in this paper in
\S \ref{s:pred}).

The free parameters in this population model are thus: $M_{\rm{mid}}$, $\gamma$,
$D_{\rm{c}}$, $e$, and $Q_{\rm{D}}^\star$.

\section{A star population model: fit to 24 $\mu$m statistics}
\label{s:rieke}
Here the debris disk population model of \S \ref{s:model} is applied
to the statistics of the incidence of disks around A stars presented
in Rieke et al. (2005) based on observations at 24 $\mu$m.
There are two main aims.
The first, which is described in the rest of this section, is to determine if these
statistics can be reproduced with steady-state evolution of the disks,
and if so to set constraints on the physical properties and starting conditions 
of the disks in that population.
The second is to use the resulting model population to make predictions for
the properties of A star disks found in surveys at other wavelengths and to
determine if this simple model can explain the trends seen in the data.
This is described in \S\S \ref{s:rsst05pred} and \ref{s:pred},
where the model parameters are also fine-tuned.

\subsection{Rieke et al. (2005) sample}
\label{ss:riekesample}
\begin{figure}
  \centering
  \begin{tabular}{c}
     \hspace{-0.0in} \includegraphics[width=3.2in]{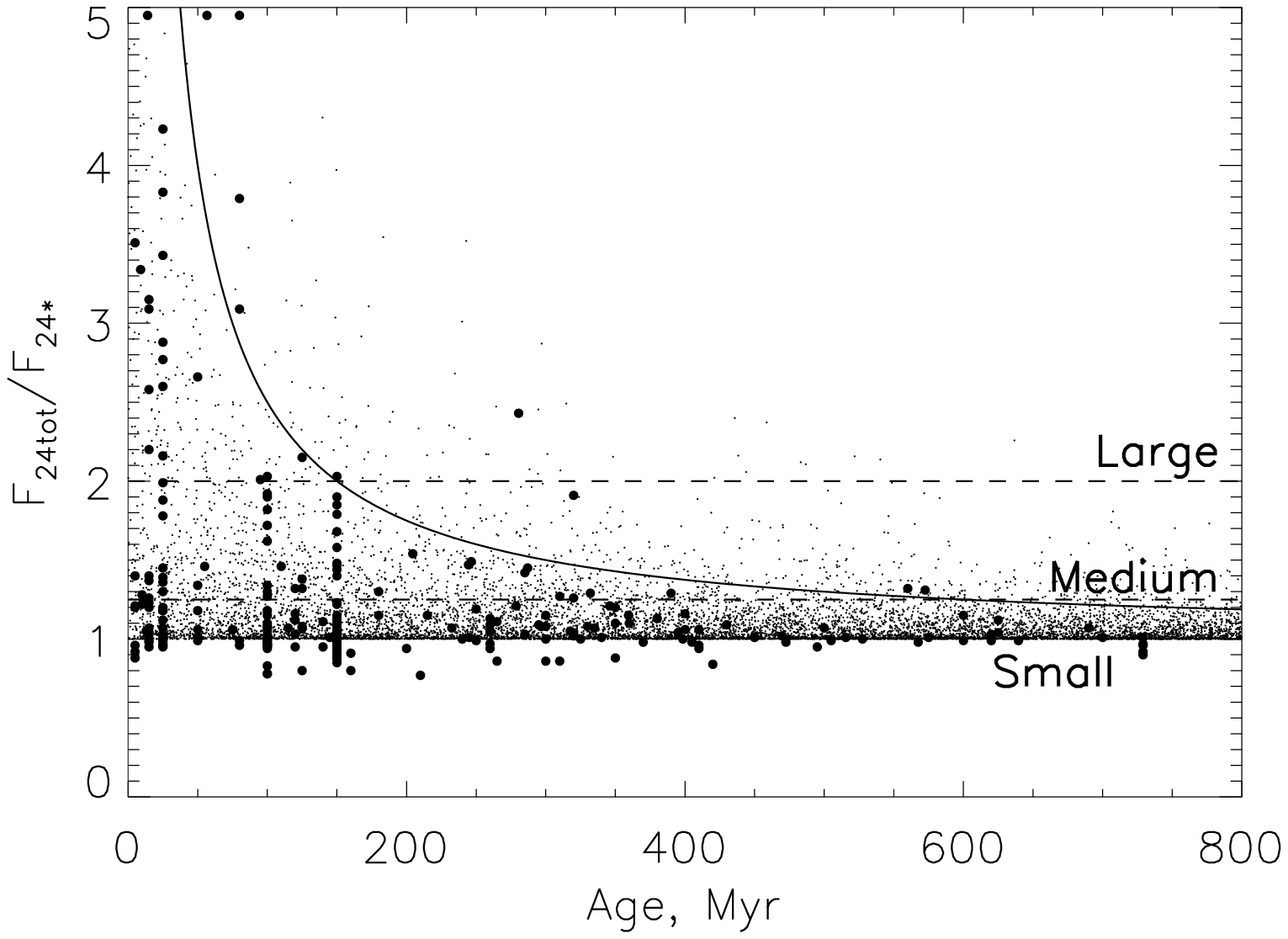} \\[-0.2in]
     \hspace{-0.0in} \includegraphics[width=3.2in]{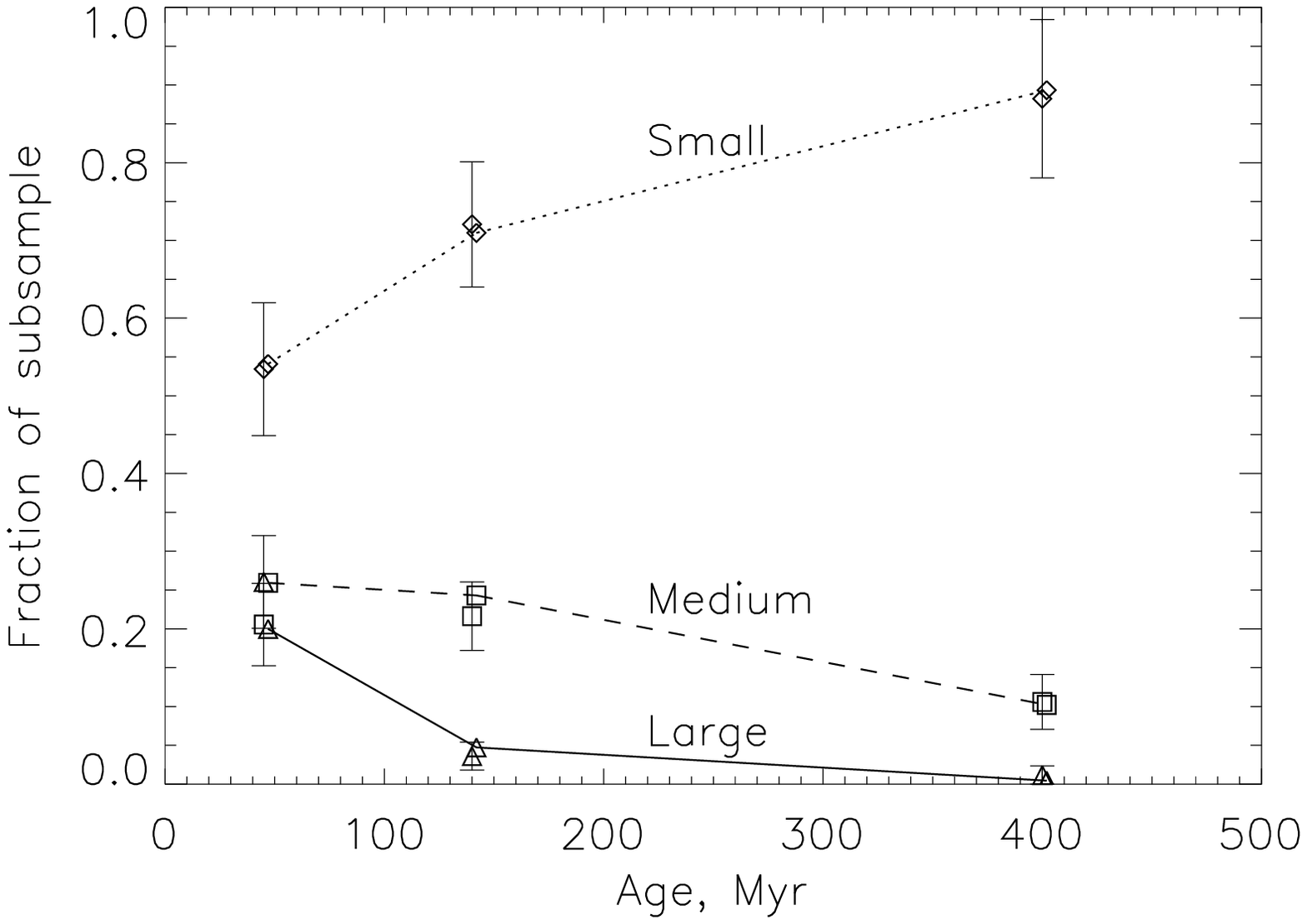} \\[-0.2in]
     \hspace{-0.0in} \includegraphics[width=3.2in]{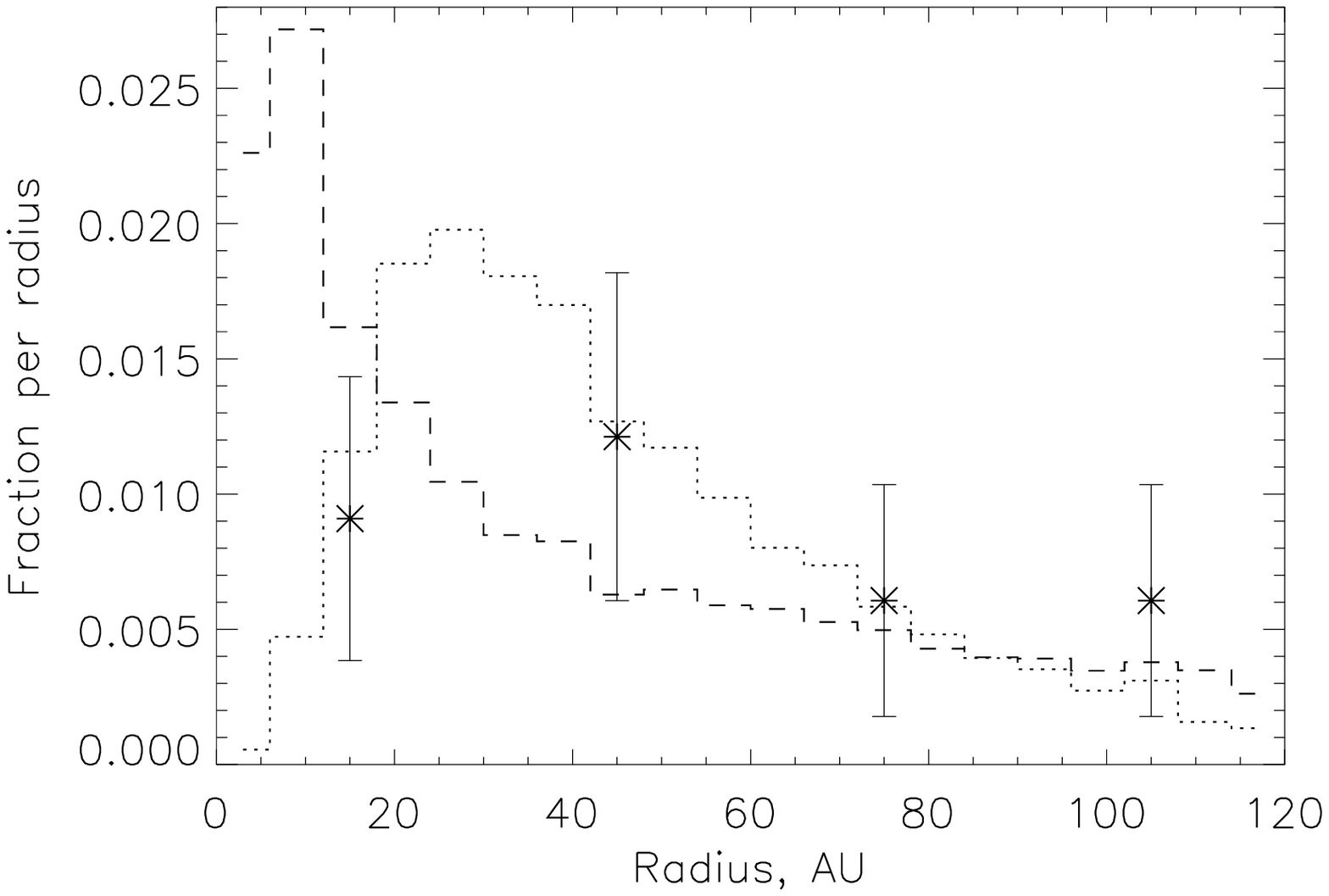} \\[-0.0in]
  \end{tabular}
  \caption{Fit of the debris disk population model to the results of
  Rieke et al. (2005) (\S \ref{s:rieke}).
  The model is comprised of 10,000 planetesimal 
  belts with initial masses chosen from a log-normal distribution centred on $10M_\oplus$, 
  radii in the range 3-120 AU (power law exponent $\gamma=-0.8$), around stars of spectral
  types B8 to A9 and ages 0-800 Myr.
  \textbf{(Top)} Total 24 $\mu$m flux divided by that of the stellar photosphere (cf. Fig. 1
  of Rieke et al. 2005).
  The model population is shown with dots, and the disk observations of Rieke et al. (2005)
  shown with filled circles.
  The upper limit inferred from statistics in Rieke et al. (2005),
  $F_{\rm{24disk}}/F_{\rm{24\star}} \approx 150t_{\rm{age}}^{-1}$, is shown with a solid line.
  \textbf{(Middle)} Fraction of stars with flux ratios in the different age bins used by Rieke et al. 
  (2005) for their Fig. 3 ($<90$ Myr, 90-190 Myr, $>190$ Myr):
  $F_{\rm{24tot}}/F_{\rm{24}\star}=$ 1-1.25 (diamond=small excess), 1.25-2 (square=medium excess), 
  $>2$ (triangle=large excess).
  Observed values are shown with $\sqrt{N}$ error bars, and model values are joined with 
  dotted, dashed and solid lines for increasing $F_{\rm{24tot}}/F_{\rm{24}\star}$.
  \textbf{(Bottom)} Distribution of radii in the model (dashed line is the whole population,
  dotted line is those predicted to be detectable at 24 and 70 $\mu$m) compared with the 
  distribution inferred from observations in Rieke et al. (2005) with $\sqrt{N}$ error bars 
  (asterisks).
  The y axis is the fraction of disks in each sample that fall in the different radius bins
  (of width 6 or 30AU).}
  \label{fig:rsst05}
\end{figure}

Rieke et al. (2005) searched a sample of 76 individual main sequence A stars
and a number of young stellar clusters for excess 24 $\mu$m emission above
photospheric levels using Spitzer and
combined the data with archival results from IRAS for a total sample size 
of 266 A stars.
They also determined the ages of all stars in a consistent manner thus
providing a very useful sample from which to study the evolution of debris
around A stars.
Their main result was a plot of $F_{\rm{24tot}}/F_{\rm{24}\star}$ versus age for all
objects which showed a decrease in the upper envelope $\sim150$ Myr$/t_{\rm{age}}$
(see Fig. \ref{fig:rsst05}).
The statistics were further quantified by splitting the sample into
three age bins ($<90$ Myr, $90-189$ Myr and $>190$ Myr), and classifying
sources as either having small excess ($F_{\rm{24tot}}/F_{\rm{24}\star}<1.25$),
medium excess ($1.25<F_{\rm{24tot}}/F_{\rm{24}\star}<2$) or large excess
($F_{\rm{24tot}}/F_{\rm{24}\star}>2$) (see their Fig. 3 and our middle Fig.~\ref{fig:rsst05}).
Data on 12 objects was also presented at 70 $\mu$m (see their Fig. 2 and
our bottom Fig.~\ref{fig:rsst05} for a different version of this data).

\subsection{Best fit model population}
\label{ss:bestfitmodel}
Our best fit model for the disks around a population of 10,000 main sequence A stars
is shown in Fig.~\ref{fig:rsst05}.
For each disk the stellar spectral type was chosen at random from the range
B8V to A9V, with the ages of the disks chosen randomly from the range 0-800 Myr
(that of the stars in the Rieke et al. sample).
All disks were assumed to have physical properties corresponding to
$Q_{\rm{D}}^\star = 300$ J kg$^{-1}$, $e=0.05$, and $D_{\rm{c}} = 60$ km.
The distribution of their initial masses had $M_{\rm{mid}}=10M_\oplus$, slightly higher
than that inferred for the disks of sun-like stars in Taurus-Auriga to account for a
slightly higher disk mass expected around more massive stars (Natta et al. 2000).
The consequence of choosing different values is discussed in \S \ref{ss:modelconstraints}.

The distribution of radii was constrained using the 12 disks with data
at 24 and 70 $\mu$m in Rieke et al. (2005).
Their values of $F_{\rm{70disk}}/F_{\rm{24disk}}$ were converted into dust
temperature and then radii (both assuming black body emission)
to derive a distribution of radii that is reasonably flat, falling off only slowly between
3 and 120 AU (bottom Fig.~\ref{fig:rsst05}).
While this sample is small this distribution compares well with the distribution of radii
inferred by Najita \& Williams (2005) from their sub-mm sample of main sequence stars,
and with that inferred from the larger sample of 46 A star disks for which radii could be 
estimated, which is discussed in \S \ref{s:rsst05pred} (see Fig.~\ref{fig:rdist}).
It is important to account for the fact that this sample of stars with estimated radii
is not representative of the whole population, since it only includes disks that can be
detected at 24 and 70 $\mu$m.
For this reason the observed distribution of radii was not compared with the
power law distribution used as input to the model, but with the sub-sample of disks in the 
model population that could have been detected at both 24 and 70 $\mu$m, which was
assumed to be those with $F_{\rm{24tot}}/F_{\rm{24}\star} > 1.1$ and
$F_{\rm{70tot}}/F_{\rm{70}\star} > 1.55$ (bottom Fig.~\ref{fig:rsst05}).
The exponent in the power law distribution was found to be $\gamma \approx -0.8 \pm 0.6$,
with relatively poor constraints due to the small sample size.

One immediately clear result is that the observed statistics can be well fitted
using the model with a realistic set of planetesimal belt parameters (i.e., those
given above) and with an initial mass distribution
consistent with that inferred for proto-planetary disks.
Thus there is no requirement for disk flux to have any stochastic component to
its evolution to explain the Rieke et al. (2005) statistics, although this does
not mean that stochasticity cannot play a role in determining the observable properties of the
disks.
The large spread in infrared excess for stars at any given age arises in the model
from the distribution of their initial starting masses, as well as from that of their
radii, and to a lesser extent from the spectral type of the parent star.
Disks of high 24 $\mu$m flux in the model are those with high initial disk masses
and with radii that are larger than average, since rapid collisional processing means 
that close in planetesimal belts tend to have low luminosities, except at the youngest 
ages.
The model also predicts that disks of high 24 $\mu$m flux should be found predominantly 
around higher luminosity stars; e.g., 40\% of the stars in the model with
$F_{\rm{24tot}}/F_{\rm{24}\star}>2$ are A0 or earlier compared with
25\% of stars in the whole population.
We have not been able to confirm this trend in the Rieke et al. (2005) data,
for which any evidence of a change in excess ratio with spectral type
is overwhelmed by the strong age dependence of this parameter.

\subsection{Constraints and flexibility in model parameters}
\label{ss:modelconstraints}
The model parameters described in \S \ref{ss:bestfitmodel} are not uniquely constrained
in the formulation of the evolution described in \S \ref{s:model}:

\textbf{(i)} The upper envelope in the model population in top Fig.~\ref{fig:rsst05} is 
determined by $f_{\rm{max}}$ given in equation (\ref{eq:fmax1}).
Thus, given the parameters and distributions that are fixed in the model,
the upper envelope seen in top Fig.~\ref{fig:rsst05} in both the model and observed
populations would be reproduced by any combination of parameters with the same
$D_{\rm{c}}^{3q-5}/G(q,X_{\rm{c}})$.
This means that if $q=11/6$ and $X_{\rm{c}} \ll 1$, then the same statistics are reproduced
as long as
\begin{equation}
  D_{\rm{c}}^{1/2}{Q_{\rm{D}}^\star}^{5/6}e^{-5/3} \approx 130 \times 10^3. \label{eq:rel1}
\end{equation}
The parameters $dr/r$, $\rho$, $e/I$ and $q$ could be changed and accounted
for by modifying the other variables in a similar manner.
The factor derived here for A stars in equation (\ref{eq:rel1}) is four times lower than
that used in Wyatt et al. (2007) to assess whether the hot dust of sun-like stars exceeds
the maximum possible for steady-state planetesimal belts, indicating that the limit used
in that paper is, if anything, too stringent (although it is possible that the 
properties of the planetesimal belts of a population of sun-like stars are different
from those of A stars given in eq.~\ref{eq:rel1}).

\textbf{(ii)} The mass distribution in the model determines the path of each disk under the
upper envelope described above.
This mass distribution thus determines the fraction of disks that end up in the
different bins of $F_{\rm{24tot}}/F_{\rm{24}\star}$ at different times.
If the relation given in equation (\ref{eq:rel1}) holds, then the same statistics
as the model shown in Fig.~\ref{fig:rsst05} would also be reproduced
with a disk mass distribution which is given by:
\begin{equation}
  M_{\rm{mid}}D_{\rm{c}}^{-0.5} \approx 1.3, \label{eq:rel2}
\end{equation}
since, from equation (\ref{eq:tcmtot}), this would result in the same distribution of
$t_{\rm{c}}$.

In other words, if a maximum planetesimal size of 2000 km had been assumed, the disk
mass distribution would have to be centred on $M_{\rm{mid}} = 60M_\oplus$, i.e., close to
an order of magnitude higher than that inferred from the distribution of 
proto-planetary disks including non-detections (Andrews \& Williams 2005), but half
the mean mass of disks detected around Herbig Ae stars (Natta et al. 2000).
In such an interpretation equation (\ref{eq:rel1}) indicates that
$e \approx 0.008 \sqrt{Q_{\rm{D}}^\star}$ and so is satisfied by
$e=0.1$ and $Q_{\rm{D}}^\star=140$ J kg$^{-1}$.

Distributions with more massive disks are possible by assuming, e.g., that some 
fraction of stars end up with only very tenuous disks (which are not detected at 24 
$\mu$m).
Lower mean disk masses are also possible to accommodate by reducing the maximum 
planetesimal size and planetesimal eccentricity;
e.g., the statistics are also reproduced with $M_{\rm{mid}} = 1M_\oplus$,
$D_{\rm{c}}=0.6$ km, $e=0.013$, $Q_{\rm{D}}^\star=300$ J kg$^{-1}$.
It is not possible to differentiate among these possibilities at this stage.
Both atypically high and low starting masses for debris disks have been suggested
by other studies:
e.g., a jump down in disk mass by a factor of $10-100$ is inferred when extrapolating
the masses of the known debris disks back in time and comparing with
proto-planetary disk masses (Wyatt, Dent \& Greaves 2003), and the statistics of the
incidence of sun-like stars with planets and debris suggest that it is stars with disks
with $\sim 20M_\oplus$ of solid material (i.e., those with masses 6 times higher than
$M_{\rm{mid}}$) that end up with detectable debris (Greaves et al. 2006).
It may also be reasonable to assume that only some fraction of the initial disk mass
ends up in the planetesimal belt, the remainder being incorporated into planets or
in some other way removed.

The radius distribution is also particularly poorly constrained at present.
On top of the large uncertainty in the power law exponent $\gamma$, it is also
possible that the range of radii for the planetesimal belts is broader
than that assumed here (3-120 AU).
The model predicts that planetesimal belts that fall outside this range would not
be detected in a 24 $\mu$m survey such as that of Rieke et al. (2005), since
the rapid collisional processing of those $<3$ AU means that these belts fall
below the detection threshold soon after they form, while those $>120$ AU would
be too cold for their emission at 24 $\mu$m to be detectable above the photosphere.
Thus we cannot rule out the existence of such close-in or far-out planetesimal
belts around up to 40\% of A stars from this study.

It is also important to point out that while the model presented here considered disks
that all had the same maximum planetesimal size but with a range of starting masses, a
model with a narrower distribution of initial masses and a distribution of maximum 
planetesimal sizes could also fit the observations.
Likewise, the model chose to set planetesimal strength, eccentricity and maximum size
to be the same for all systems.
It is not a conclusion that this has to be the case, since a population for which
these parameters are chosen from distributions can also fit the data.
In that instance the properties of the planetesimal belts would also contribute
to the spread in infrared excess seen at any given age.

In conclusion the statistics can be readily reproduced with a population of
disks with reasonable starting parameters without recourse to stochastic
evolution models.
The possibility remains, however, that some fraction of the disks in the
Rieke et al. (2005) sample are undergoing transient events.
For example, some of the disks may exceed the maximum luminosity permitted for disks
at their observed radius and age.
This limit is given in equation (\ref{eq:fmax1}) and was used in Wyatt et al. (2007) to show
that several of the disks found $<10$ AU around sun-like stars must be transient.
One system in the Rieke et al. (2005) sample that is close to this limit is $\zeta$ Lep
(HD38678), a 230 Myr A2V star which has dust with an infrared luminosity
$f=10\times 10^{-5}$ (Chen \& Jura 2001; Su et al. 2006) predicted to lie at
$\sim 7$ AU based on its 24 and 70 $\mu$m fluxes, and found to have a resolved
size of $\sim 3$ AU at 18 $\mu$m, with emission spread across the range
2-8 AU (Moerchen et al. 2007).
For dust at 7 AU $f_{\rm{max}}$ is $0.16 \times 10^{-5}$ using the parameters
inferred above for the rest of the A star sample (see also Table \ref{tab:2470sample});
i.e., this system is close to the limit at which we would infer it must be transient which was
considered in Wyatt et al. (2007) to be when $f > 100f_{\rm{max}}$.
Further some fraction of the disks could have a two component structure, e.g., like the
F2V star $\eta$ Corvi which has dust at $\sim 1.5$ AU that is inferred to be transient
(Wyatt et al. 2007), as well as at 150 AU (Wyatt et al. 2005), which is likely to be
in steady-state.
Indeed it has recently been shown that Vega (HD172167)
does have an additional dust component at a few AU (Absil et al. 2006), as was already
inferred for HR4796 (HD109573; Augereau et al. 1999).
Also, there could be a stochastic element to the mid-IR emission which is not present
in the far-IR emission, such as has been inferred for Vega based on the fact that the
mid-IR emission seems to be dominated by grains undergoing blow-out by radiation
pressure (Su et al. 2005).
Detailed study of individual objects is required to resolve these uncertainties
on a case-by-case basis.
Since the threshold for detection of a stochastic event is high with the current
observations, it is likely that the number of such cases is underestimated.
Nonetheless, the correspondence of the time decay of the excesses with a simple model
of steady-state evolution that has few free parameters suggests that stochastic events
do not dominate the behavior.

\subsection{Interpretation of model parameters}
\label{ss:interpretation}
Given the degeneracy in the model parameters described in \S \ref{ss:modelconstraints}
it is not worth dwelling on a specific interpretation of the values used in the best
fit model of \S \ref{ss:bestfitmodel}, except to say that these are reasonable
parameters based on models of planetesimal strength (Benz \& Asphaug 1999)
and of planet formation processes (Kenyon \& Bromley 2002).
A more detailed collisional evolution model would be required to ascertain what this
is telling us about the physical properties of the planetesimals, for example.
The important point is that the observed statistics can be explained
by a model population in which the evolution of individual disks is very simple
and behaves in the manner described in \S \ref{s:model}.

\section{Explaining 24 and 70 $\mu$m A star results}
\label{s:rsst05pred}
\begin{figure*}
  \centering
  \begin{tabular}{cc}
     \hspace{-0.2in} \includegraphics[width=3.2in]{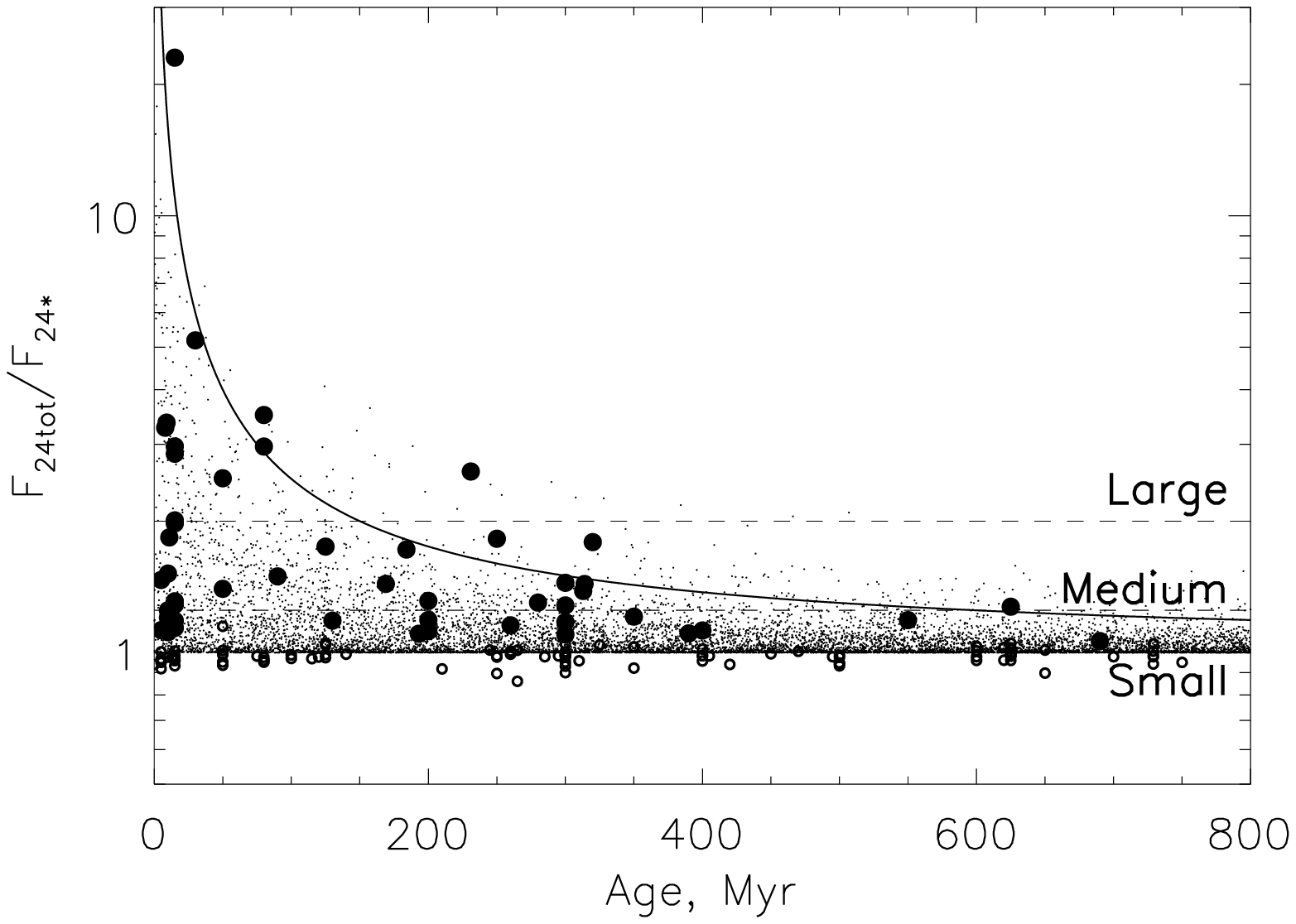} &
     \hspace{-0.2in} \includegraphics[width=3.2in]{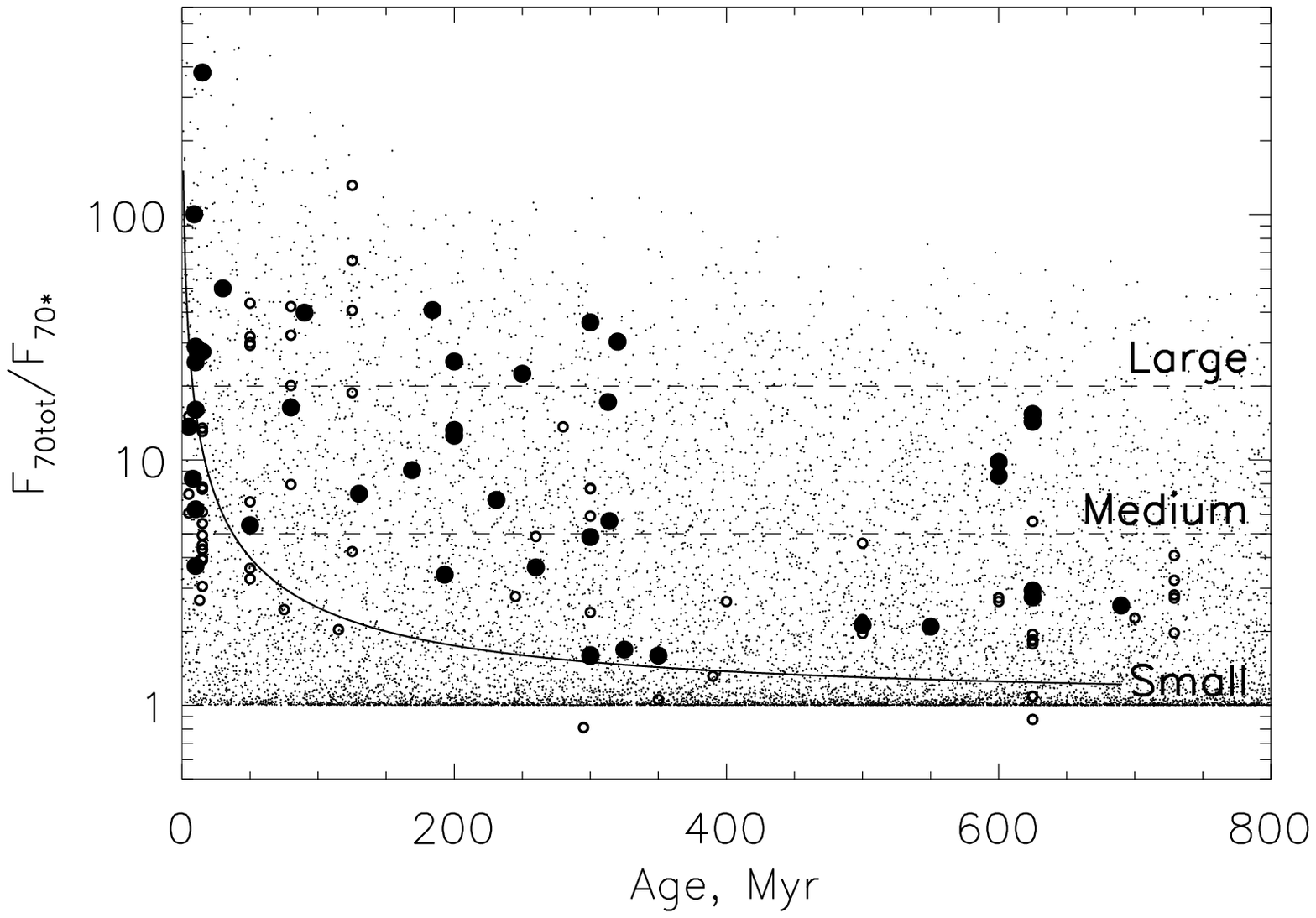} \\[-0.2in]
     \hspace{-0.2in} \includegraphics[width=3.2in]{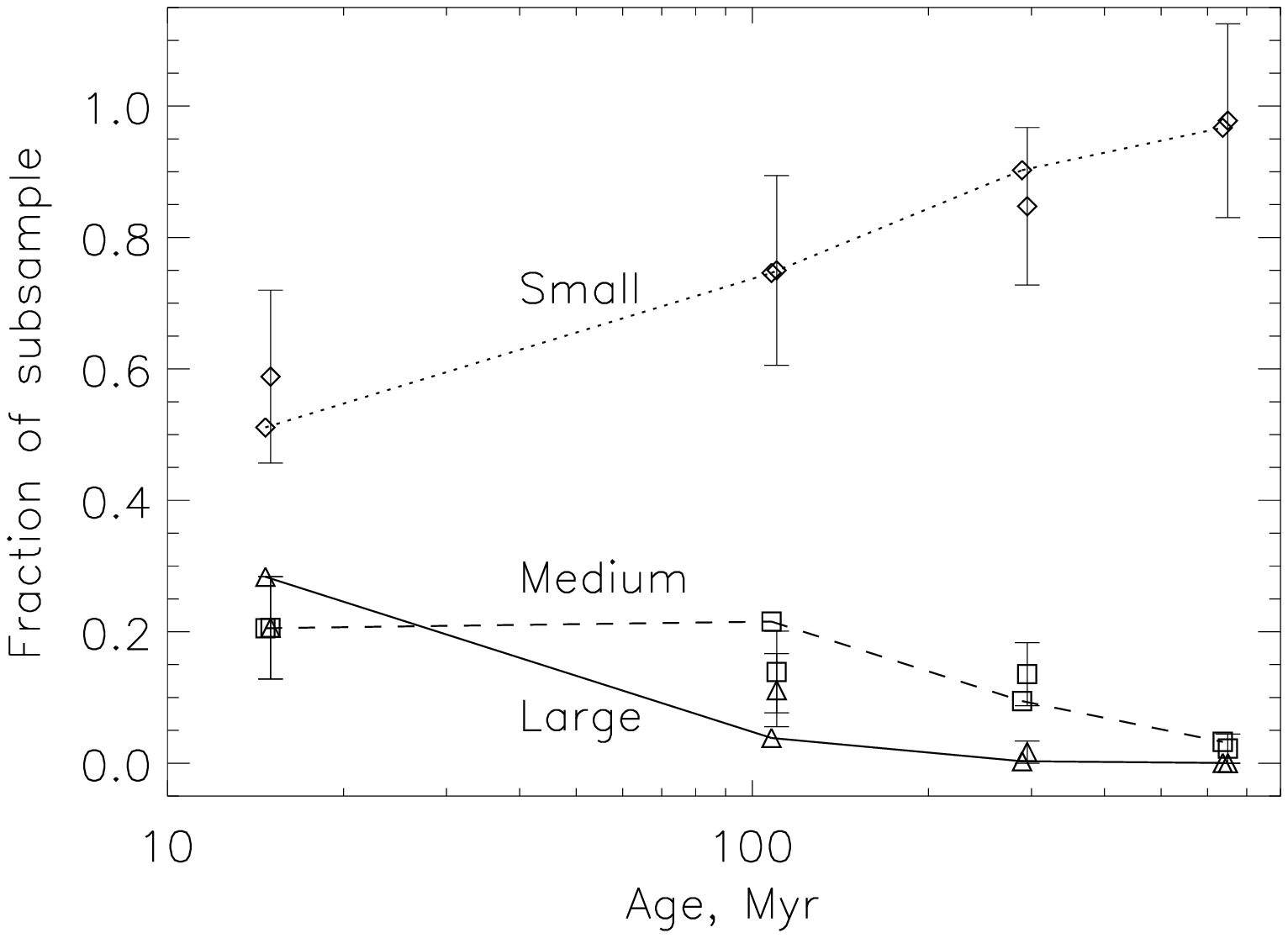} &
     \hspace{-0.2in} \includegraphics[width=3.2in]{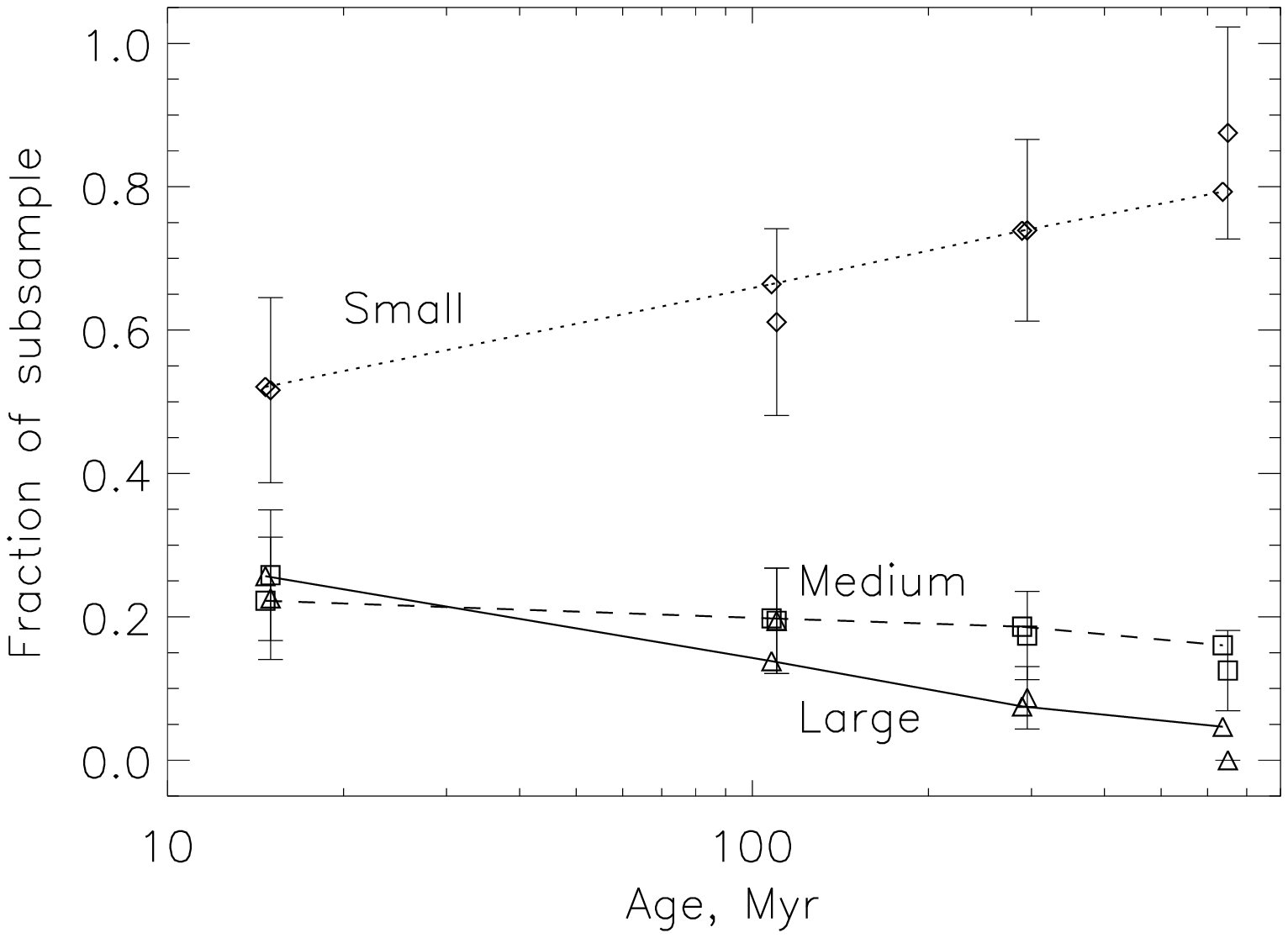} \\[-0.0in]
  \end{tabular}
  \caption{Fit of the model population to the distributions of: \textbf{(Left)} 24 $\mu$m and
  \textbf{(Right)} 70 $\mu$m excesses in Su et al. (2006) (\S \ref{s:rsst05pred}).
  The model parameters are unchanged to those of \S \ref{s:rieke} (Fig.~\ref{fig:rsst05})
  except for a small change in planetesimal strength.
  \textbf{(Top)} Total fluxes at 24 and 70 $\mu$m flux divided by that of the stellar photosphere
  (cf. Figs. 4 and 5 of Su et al. 2006).
  The 10,000 disks in the model population are shown with dots.
  Stars with excess emission observed at 24 and 70 $\mu$m are shown with filled circles in the
  left and right plots respectively, while those undetected at 24 $\mu$m, or for which
  upper limits were derived at 70 $\mu$m, are shown with open circles.
  The $150t_{\rm{age}}^{-1}$ line delineating the upper envelope of the 24 $\mu$m detections 
  (Rieke et al. 2005) is shown with a solid line.
  \textbf{(Bottom)} Fraction of stars with flux ratios in the different age bins used by Su et 
  al. (2006) for their Fig. 9 ($<30$ Myr, 30-190 Myr, 190-400 Myr, $>400$ Myr):
  $F_{\rm{24tot}}/F_{\rm{24}\star}=$ 1-1.25 (diamond=small excess), 1.25-2 (square=medium excess),
  $>2$ (triangle=large excess);
  and $F_{\rm{70tot}}/F_{\rm{70}\star}=$ 1-5 (diamond=small excess), 5-20 (square=medium excess), $>20$ 
  (triangle=large excess).
  Observed values are shown with $\sqrt{N}$ error bars, and model values are joined with 
  dotted, dash and solid lines for increasing $F_{\rm{tot}}/F_\star$.
  }
  \label{fig:70um}
\end{figure*}

While there is still uncertainty in how accurately the simple collisional
evolution model of \S \ref{ss:stst} follows the evolution of individual disks 
(e.g., the validity of a single power law size distribution; see
Wyatt et al. 2007), the results of \S \ref{s:rieke} 
show that the debris disk population model based on that evolution
can reproduce the 24 $\mu$m statistics well.
In this section, the model population is further tested by comparing its
predictions for the surveys at 70 $\mu$m with the statistics observed by
Su et al. (2006) (\S \ref{ss:70um}).
The model parameters are also fine-tuned in this section to provide a better fit
to the observed statistics from that paper.
By plotting the model population in different ways we are able to
provide a qualitative explanation of the trends seen in this survey,
and to reproduce the properties of a sample of 46 A stars for which
excess emission has been detected at both 24 and 70 $\mu$m and so for
which estimates for the radii of their planetesimal belts can be made (\S \ref{ss:2470}).
These plots provide a framework that can be used to interpret the
statistics of surveys at other wavelengths, and to understand the
implications of different detection thresholds.

\subsection{Fit to Su et al. (2006) 70 $\mu$m statistics}
\label{ss:70um}

The Rieke et al. (2005) 24 $\mu$m survey of A stars for excess emission described in
\S \ref{ss:riekesample} has recently been extended to 70 $\mu$m and the results presented
in Su et al. (2006).
In that paper a sample of $\sim 160$ main sequence A-type stars (spectral types B6 to A7)
were searched for excesses at both 24 and 70 $\mu$m using Spitzer, and ages were assigned
to these stars using a consistent scheme.
This sample was then supplemented with a further 19 stars with IRAS data at 25 and 60 $\mu$m.
In addition to reproducing the results for the evolution of 24 $\mu$m excesses around
A stars found by Rieke et al. (2005) (e.g., Fig.~\ref{fig:rsst05}), the main new result
of Su et al. (2006) was comparable plots for the evolution of 70 $\mu$m 
excesses.
Plots of $F_{\rm{tot}}/F_{\rm{\star}}$ against $t_{\rm{age}}$ for both 24 and
70 $\mu$m (Figs. 4 and 5 of Su et al. 2006) and the fraction of stars in 
different age bins that have
small excess ($F_{\rm{24tot}}/F_{\rm{24\star}}<1.25$, $F_{\rm{70tot}}/F_{\rm{70\star}}<5$),
medium excess ($1.25<F_{\rm{24tot}}/F_{\rm{24\star}}<2$,$5<F_{\rm{70tot}}/F_{\rm{70\star}}<20$),
and large excess ($F_{\rm{24tot}}/F_{\rm{24\star}}>2$, $F_{\rm{70tot}}/F_{\rm{70\star}}>20$)
(Fig. 9 of Su et al. 2006) are shown in Fig.~\ref{fig:70um}.
The observations show that while 70 $\mu$m excesses are also characterized by a
$t_{\rm{age}}^{-1}$ fall-off, the disk emission persists much longer, and at much
higher levels relative to the photospheric flux, than at 24 $\mu$m.
Su et al. (2006) also identified samples of stars for which excess emission was 
detected at both 24 and
70 $\mu$m (group I sources in Su et al.), or at 70 but not 24 $\mu$m (group II sources in
Su et al.), or at 24 but not 70 $\mu$m (group V sources in Su et al.).

\begin{figure}
  \centering
  \begin{tabular}{c}
     \hspace{-0.0in} \includegraphics[width=3.2in]{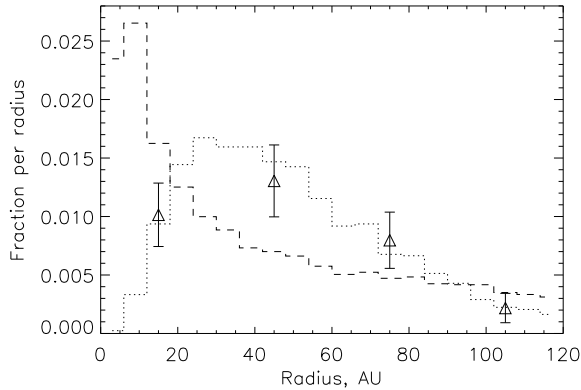}
  \end{tabular}
  \caption{
  Distribution of planetesimal belt radii.
  That of the whole model population described in \S \ref{s:rsst05pred} (and shown in Fig.~\ref{fig:70um})
  is shown with a dashed line, while that of the sub-sample of the model population predicted to be detectable
  at 24 and 70 $\mu$m is shown with a dotted line.
  The observed distribution of radii for the 46 A star disks that have been
  detected at both 24 and 70 $\mu$m (\S \ref{ss:2470sample})
  is shown with triangles and $\sqrt{N}$ error bars.
  The y axis is the fraction of disks in each sample that fall in the different radius bins
  (of width 6 or 30AU).}
  \label{fig:rdist}
\end{figure}

The model of \S \ref{s:rieke} was fitted to the distribution of disk
properties above the 24 $\mu$m threshold, and the only way 70 $\mu$m information
was included in the model fit was through the 12 disks with 70 $\mu$m detections
in Rieke et al. (2005) which set rather loose constraints on the exponent in the
power law distribution of planetesimal belt radii (bottom Fig.~\ref{fig:rsst05}).
In fact it turns out that the 70 $\mu$m statistics of Su et al. (2006) are 
fitted
extremely well using exactly the same model population.
However, we used the new distributions of 24 and 70 $\mu$m excesses as a function of
age shown in Fig.~\ref{fig:70um}, and the distribution of radii for the
46 A star disks that have been detected at 24 and 70 $\mu$m (see \S \ref{ss:2470} and
Figure \ref{fig:rdist}), to set new constraints on the model parameters.
Figures (\ref{fig:70um}) and (\ref{fig:rdist}) show the resulting best fit model
population, and illustrate how the model reproduces the longer timescale for the
decay of 70 $\mu$m excesses seen by Su et al. (2006), and not just in a
qualitative way:
the model reproduces the observed statistics very well at 70 $\mu$m, just as it does
at 24 $\mu$m.
While the upper envelope in top right Fig.~\ref{fig:70um} is around 5 times as high as that
of the observed disks at any given age, this can be attributed to small number statistics,
since the observed upper envelope would increase as more stars are included in the sample,
and indeed the upper envelope in the model population is close to that
observed when the number of stars included is comparable with the observed population
(i.e., two orders of magnitude fewer stars in the model population than shown
in Fig.~\ref{fig:70um}).

The parameters derived from this fit are almost completely unchanged from 
those derived in \S \ref{s:rieke}, with the only change being in the planetesimal
strength to ${Q_{\rm{D}}}^\star = 150$ J kg$^{-1}$.
This means that the statistics can equally be fitted by any combination of parameters:
\begin{equation}
  D_{\rm{c}}^{1/2}{Q_{\rm{D}}^\star}^{5/6}e^{-5/3} \approx 74 \times 10^3, \label{eq:rel1v2}
\end{equation}
which now supersedes equation (\ref{eq:rel1}), while the combination of parameters required to
fit the statistics that is given in equation (\ref{eq:rel2}) remains unchanged.
The exponent in the radius distribution also remains unchanged as a result of the new fit,
but this distribution is now better constrained to $N(r) \propto r^{-0.8 \pm 0.3}$.

\subsection{Comparison with disks detected at 24 and 70 $\mu$m}
\label{ss:2470}
In addition to the quantitative results on the incidence of disks of different levels
of excess seen around stars of different ages, Su et al. (2006) identified 
several
trends in their data which also require explanation in terms of the model.
These include:
(i) disks with excess at 70 $\mu$m but not 24 $\mu$m have lower fractional luminosities
than those detected at both 24 and 70 $\mu$m;
(ii) stars that have excesses at 70 $\mu$m but not at 24 $\mu$m are on average older than
the rest of the population;
\footnote{HD188228 in Su et al's group II has an age of $\sim 10$ Myr and is an
exception to the rule they state that young stars with 70 $\mu$m excesses also have
an excess at 24 $\mu$m.
However, the remaining 7 stars in their group II have ages $>325$ Myr, and the
mean age of their group II stars, $\sim 490$ Myr, is significantly older
than the rest of the population.}
(iii) stars with no apparent excess have limits as low as $f<10^{-7}$;
(iv) a large spread of two orders of magnitude of fractional luminosity at any given age;
(v) older stars have lower fractional luminosity than younger ones;
(vi) the upper envelope in fractional luminosity falls off $\propto t_{\rm{age}}^{-1}$,
and is not constant as found in Decin et al. (2003);
(vii) for disks detected at both 24 and 70 $\mu$m, the color temperature distribution
is broader at younger ages.

Here we explain these trends by considering the effect of the detection threshold on
the sub-samples of the model population that are detected at 24 or 70 $\mu$m, or at
both wavelengths, on 3 plots of the model population.
The model is compared with a sample of 46 A stars for which excess emission was detected
at both 24 and 70 $\mu$m and so for which the radii of their
disks can be estimated;
that sample is described in \S \ref{ss:2470sample}.
The three plots are the three fundamental disk parameters plotted against
each other: $f$ versus $r$ (\S \ref{ss:fvsr}),
$f$ versus $t_{\rm{age}}$ (\S \ref{ss:fvst}),
and $r$ versus $t_{\rm{age}}$ (\S \ref{ss:rvst}).
A further comparison of the model to the 24 and 70 $\mu$m detected sample is
given in \S \ref{ss:fmax}.

\subsubsection{Sample of 46 A star disks detected at 24 and 70 $\mu$m}
\label{ss:2470sample}

\begin{deluxetable*}{ccccccccccc}
  \tabletypesize{\scriptsize}
  \tablecaption{Main sequence A-type stars with excess emission detected at 24 and 70 $\mu$m (or 25 and 60 $\mu$m)
  \label{tab:2470sample} }
  \tablewidth{0pt}
  \tablehead{
    \colhead{Star name} &
    \colhead{Sp. Type} &
    \colhead{$T_\star$, K} &
    \colhead{$M_\star$, $M_\sun$} &
    \colhead{$L_\star$, $L_\sun$} &
    \colhead{$d$, pc} &
    \colhead{$t_{\rm{age}}$, Myr} &
    \colhead{$f = L_{\rm{ir}} / L_\star $} &
    \colhead{$r$, AU} &
    \colhead{ $f/f_{\rm{max}}$ } & 
    \colhead{ $f/f_{\rm{pr}}$ } }
  \startdata
\multicolumn{11}{l}{Su et al. (2006) sources} \\
HD2262 & A7V &   9506&    2.9&   12.0&   23.5 &          690 & $    0.71 \times 10^{-5}$ &           28 &   1.0 &      \textbf{0.9} \\
HD14055 & A1Vnn &   9225&    2.7&   33.0&   36.1 &          300 & $    6.7 \times 10^{-5}$ &           79 &   0.65 &    15 \\
HD19356 & B8V &  11939&    3.8&  347&   28.5 &          300 & $    0.32 \times 10^{-5}$ &           30 &   1.1 &     \textbf{0.4} \\
HD27045 & A3m &   8709&    2.3&    8.0&   28.7 &          193 & $    1.1 \times 10^{-5}$ &           22 &   0.49 &     1.3 \\
HD28355 & A7V &   7852&    1.8&   18.0&   49.2 &          625 & $    5.6 \times 10^{-5}$ &           49 &   1.8 &      12 \\
HD30422 & A3IV &   8709&    2.3&    9.0&   57.5 &           10 & $    4.9 \times 10^{-5}$ &           40 &   0.033 &    8.1 \\
HD31295 & A0V &   9506&    2.9&   26.0&   37.0 &           10 & $    4.5 \times 10^{-5}$ &           78 &   0.014 &     9.3 \\
HD38056 & A0V &   9506&    2.9&   54.0&  132 &          250 & $    4.8 \times 10^{-5}$ &           62 &   0.91 &     8.9 \\
HD38206 & A0V &   9506&    2.9&   32.0&   69.2 &            9 & $     14 \times 10^{-5}$ &           59 &   0.078 &      25 \\
HD38678 & A2Vann &   8974&    2.5&   16.0&   21.5 &          231 & $    9.8 \times 10^{-5}$ &            7 &  \textbf{100} &     6.7 \\
HD39060 & A5V &   8203&    2.0&   13.0&   19.3 &           15 & $     140 \times 10^{-5}$ &           24 &   5.0 &     200 \\
HD71043 & A0V &   9506&    2.9&   31.0&   73.1 &           11 & $    4.7 \times 10^{-5}$ &           52 &   0.044 &     8.0 \\
HD71155 & A0V &   9506&    2.9&   40.0&   38.3 &          169 & $    2.5 \times 10^{-5}$ &           44 &   0.58 &     3.9 \\
HD75416 & B8V &  11939&    3.8&  111&   96.9 &            8 & $    4.7 \times 10^{-5}$ &           17 &   0.96 &     4.0 \\
HD79108 & A0V &   9506&    2.9&   60.0&  115 &          320 & $    5.0 \times 10^{-5}$ &           77 &   0.76 &      10 \\
HD80950 & A0V &   9506&    2.9&   39.0&   80.8 &           80 & $    7.5 \times 10^{-5}$ &           20 &   4.7 &     7.9 \\
HD95418 & A1V &   9225&    2.7&   68.0&   24.3 &          300 & $    1.3 \times 10^{-5}$ &           47 &   0.55 &     2.1 \\
HD97633 & A2V &   8974&    2.5&  141&   54.5 &          550 & $    0.71 \times 10^{-5}$ &           36 &   1.4 &     1.1 \\
HD102647 & A3V &   8709&    2.3&   17.0&   11.1 &           50 & $    2.0 \times 10^{-5}$ &           21 &   0.40 &     2.4 \\
HD106591 & A3V &   8709&    2.3&   26.0&   25.0 &          300 & $    0.50 \times 10^{-5}$ &           16 &   1.5 &     \textbf{0.5} \\
HD110411 & A0V &   9506&    2.9&   23.0&   36.9 &           10 & $    3.7 \times 10^{-5}$ &           53 &   0.026 &     6.3 \\
HD111786 & A0III &   9506&    2.9&   22.0&   60.2 &          200 & $    3.0 \times 10^{-5}$ &           47 &   0.53 &     4.7 \\
HD115892 & A2V &   8974&    2.5&   26.0&   18.0 &          350 & $    1.1 \times 10^{-5}$ &            6 &  \textbf{32} &   \textbf{0.7} \\
HD125162 & A0p &   9506&    2.9&   16.0&   29.8 &          313 & $    5.1 \times 10^{-5}$ &           42 &   1.6 &     7.7 \\
HD136246 & A1V &   9225&    2.7&   31.0&  144 &           15 & $    4.8 \times 10^{-5}$ &           81 &   0.021 &      11 \\
HD139006 & A0V &   9506&    2.9&   83.0&   22.9 &          314 & $    1.4 \times 10^{-5}$ &           45 &   0.85 &     2.3 \\
HD158460 & A1V &   9225&    2.7&  105&  104 &          260 & $    0.53 \times 10^{-5}$ &           68 &   0.11 &     1.1 \\
HD161868 & A0V &   9506&    2.9&   29.0&   29.1 &          184 & $    7.5 \times 10^{-5}$ &           64 &   0.72 &      14 \\
HD165459 & A2V &   8974&    2.5&   13.0&   89.3 &            5 & $    4.7 \times 10^{-5}$ &           31 &   0.037 &     6.6 \\
HD172167 & A0V &   9506&    2.9&   58.0&    7.8 &          200 & $    2.3 \times 10^{-5}$ &          114 &   0.093 &     5.8 \\
HD181296 & A0Vn &   9506&    2.9&   22.0&   47.7 &           30 & $   20 \times 10^{-5}$ &           25 &   2.1 &     23 \\
HD183324 & A0V &   9506&    2.9&   22.0&   59.0 &           10 & $    1.0 \times 10^{-5}$ &           51 &   0.008 &     1.8 \\
HD216956 & A3V &   8709&    2.3&   18.0&    7.7 &          200 & $    6.1 \times 10^{-5}$ &           67 &   0.38 &      13 \\
HD221756 & A1III &   9225&    2.7&   27.0&   71.6 &          130 & $    1.9 \times 10^{-5}$ &           50 &   0.21 &   3.3 \\
HD225200 & A0V &   9506&    2.9&   47.0&  129 &           90 & $    8.0 \times 10^{-5}$ &           94 &   0.20 &      18 \\
\hline
\multicolumn{11}{l}{Additional sources from literature} \\
HD1438\tablenotemark{a} & B8V &  11939&    3.8&  219&  212 &           95 & $   26 \times 10^{-5}$ &          114 &   1.1 &    57 \\
HD3003\tablenotemark{b} & B8V &   9500&    2.9&   21.0&   46.5 &           50 & $   11 \times 10^{-5}$ &            7 &     \textbf{32} &     6.7 \\
HD9672\tablenotemark{a} & A1V &   9225&    2.7&   22.8&   61.3 &           20 & $   72 \times 10^{-5}$ &           68 &   0.54 &   140 \\
HD21997\tablenotemark{b} & A3IV/V &   8750&    1.9&   14.4&   73.8 &          50 & $   47 \times 10^{-5}$ &           79 &   0.39 &   120 \\
HD109573\tablenotemark{b} & A0V &  10000&    2.9&   24.3&   67.1 &            8 & $  330 \times 10^{-5}$ &           27 &   8.7 &   410 \\
HD110058\tablenotemark{a} & A0V &   9506&    2.9&   10.0&   99.9 &           10 & $  140 \times 10^{-5}$ &           18 &   7.5 &   140 \\
HD141569\tablenotemark{b} & B9.5e &  10500&    3.1&   24.2&   99.0 &            5 & $  460 \times 10^{-5}$ &           35 &   4.4 &   620 \\
HD153053\tablenotemark{b} & A5IV/V &   8000&    1.8&   12.3&   50.7 &          420 & $    7.6 \times 10^{-5}$ &           49 &   1.4 &    16 \\
HD158352\tablenotemark{b} & A8V &   7750&    1.7&   24.0&   63.1 &          600 & $    7.8 \times 10^{-5}$ &           82 &   0.85 &    22 \\
HD172555\tablenotemark{b} & A5IV/V &   8000&    2.0&    9.5&   29.2 &           12 & $   51 \times 10^{-5}$ &            4 &  \textbf{86} &    28 \\
HD176638\tablenotemark{a} & B9.5V &  10000&    3.1&   47.3&   56.3 &          200 & $    7.7 \times 10^{-5}$ &           41 &   2.9 &    11 \\

\enddata
\tablenotetext{a}{Based on excess emission detected at 25 and 60 $\mu$m by IRAS}
\tablenotetext{b}{Based on excess emission detected at 24 and 70 $\mu$m by Spitzer}
\end{deluxetable*}

In the absence of large numbers of resolved images of disks from which to
get a direct measure of their radii, the only way to estimate this
important parameter is from interpretation of the spectrum of the excess
emission, which must be detected at more than one wavelength to allow an
estimate, rather than a limit, to be determined.
Here we construct a sample of A star disks for which radius estimates
can be made since they were all detected at both 24 and 70 $\mu$m.

The sample of 46 stars is given in Table \ref{tab:2470sample} and includes:

\begin{itemize}

\item The 35 group I sources in Table 3 of Su et al. (2006) that
have excesses detected at both 24 and 70 $\mu$m (i.e., excluding HD23862
for which there is evidence of gas emission lines in its mid-IR spectrum taken with
IRS).
The fractional luminosities of these sources were taken from Table 3 of Su et 
al. (2006),
and the 24-70 $\mu$m color temperature combined with knowledge of the stellar luminosity
to infer a planetesimal belt radius from equation (\ref{eq:tbb}), i.e., by assuming this
arises from black body emission.
Stellar ages, distances and spectral types are from Table 1 in Su et al. (2006),
and stellar temperatures and masses were estimated from spectral type.

\item Eleven additional stars from the literature for which excess emission
has been detected at both 25 and 60 $\mu$m by IRAS.
Spitzer observations at 24 and 70 $\mu$m for seven of these are available either
in the literature or in the archive, and the available observations for these sources
were analyzed in the same way as the Su et al. (2006) sources to derive stellar
luminosity, dust luminosity and dust temperature (and so planetesimal belt radius).
For the remaining four sources, their 25 and 60 $\mu$m fluxes were taken from IRAS SCANPI
\footnote{http://scanpi.ipac.caltech.edu:9000}, and these fluxes were analyzed using
a process analogous to that used in Su et al. (2006).
In that process, spectral types were taken from SIMBAD and used to determine
stellar temperatures and masses, and distances were taken from Hipparcos.
Stellar luminosities were determined by integrating under the spectrum
of a Kurucz model atmosphere fitted to the K band flux from
2MASS.
The stellar spectrum was also used to estimate the photospheric contribution to the
25 and 60 $\mu$m IRAS fluxes (having taken the appropriate color correction into
account) and so to determine the temperature, radius and fractional luminosity of the
excess emission.
Fractional luminosities and radii are consistent for the 3 sources that overlap with
Rhee et al. (2007) who performed similar calculations.
The ages of these eleven sources were taken from the literature (HD1438 from Wyatt
et al. 2003; HD3003 from Song et al. 2001; HD9672, HD21997, HD110058, HD158352,
HD172555, and HD176638 from Rhee et al. 2007; HD109573 from Stauffer et al. 
1995;
HD141569 from Weinberger et al. 2000; HD153053 from Chen et al. 2006).

\end{itemize}

For consistency we checked that the parameters given in Table \ref{tab:2470sample}
can be used to predict the observed stellar and excess fluxes using the equations
given in this paper.
The stellar flux should be given by equation (\ref{eq:fnustar}), while the excess flux
can be derived by rearranging equations (\ref{eq:f}), (\ref{eq:fnu}) and (\ref{eq:tbb}) to get 
the disk flux in Jy at a wavelength $\lambda$ to be:
\begin{equation}
  F_{\nu \rm{disk}} = 2.95 \times 10^{-10} f r^2 d^{-2} B_\nu(\lambda,278.3L_\star^{0.25}r^{-0.5}).
  \label{eq:fnudisk2}
\end{equation}
Equation (\ref{eq:fnudisk2}) predicts the observed disk fluxes
at 24 and 70 $\mu$m to within 2\%, which is to be expected, since these fluxes were
used to derive the disk parameters.
\footnote{The observed disk flux here is the total observed flux less the photospheric
contribution.
As such there is an uncertainty in the true level of disk flux due to observational
uncertainties and the accuracy of the photospheric subtraction.}
At wavelengths other than 24 and 70 $\mu$m, there may be inaccuracies due to the assumption
that the emission spectrum is described by that of a black body at a single temperature
(e.g., see \S \ref{s:pred}).
The resulting stellar fluxes are also in general agreement, although they are for some
stars in error by a factor of up to 2.
We attribute this to the shape of the stellar spectrum, which means that
a bolometric luminosity cannot be rigidly applied to derive stellar fluxes to a greater
accuracy, but consider that this accuracy is sufficient to perform a simple analysis
of this sample.

Since some of these disks have been resolved in ground- and space-based imaging, it is worth 
comparing the resolved sizes with those predicted by this crude analysis of the 24 and 70 $\mu$m
fluxes.
For the disks seen to be confined to rings it is clear that the prediction
given in Table \ref{tab:2470sample} underestimates the true size by a factor of $2-3$:
Fomalhaut is predicted to have dust at 67 AU whereas 133 AU is its observed radius
(Kalas et al. 2005), and HR4796 is predicted to have dust at 27 AU whereas 70 AU is
observed (Telesco et al. 2000).
Likewise HD181327 is reported to have an observed size that is
3 times larger than that predicted using black body grains (Schneider et al. 2006).
This discrepancy can be readily explained by the fact that the smallest dust grains
in their distributions, which dominate the cross-sectional area, emit at a higher
temperature than black body for a given distance from the star (Wyatt et al. 1999;
Wyatt \& Dent 2002).
For the two young systems HD39060 ($\beta$ Pictoris) and HD141569, the dust is observed to
span a large range of radii: from tens of AU out to 1000s of AU for $\beta$ Pictoris
with a density distribution that peaks at 75 AU
(Kalas \& Jewitt 1995; Telesco et al. 2005), and from $\sim 10$ to 1200 AU for HD141569
(Fisher et al. 2000; Clampin et al. 2003).
These extended distributions certainly complicate the interpretation of these systems in
terms of rings, although it is believed that radiation forces cause the distributions
of dust in these systems to extend far beyond those of the planetesimal belts that 
created them (Augereau et al. 2001), so that the predicted planetesimal belt radii of
24 AU for $\beta$ Pictoris and 35 AU for HD141569 are not as inaccurate as they first
appear.
For Vega, the predicted size of 114 AU is close to the size observed in the
sub-mm (Holland et al. 1998), even if the far-infrared emission on which the
size estimate is based extends out to much larger distances (Su et al. 2005).
We have opted to retain the radius derived from the SED modeling for
plotting purposes, since these are then derived in a more consistent manner
across the dataset.

All but four of the disks in this sample were detected by Spitzer at both 24 and 70 $\mu$m.
The remaining four disks have yet to be observed by Spitzer, but were detected by IRAS
at both 25 and 60 $\mu$m.
Thus we consider that the sample derived here is representative of those disks which it
is possible to detect at 24 and 70 $\mu$m, and in the following sections we compare this 
sample with the sub-sample of the model population that would have been
detected in the study of Su et al. (2006), assuming the detection limits quoted 
by
those authors.
The limiting factor for these observations was not instrumental sensitivity,
rather it was the ability to distinguish between photospheric and excess emission.
It was thus the accuracy of the extrapolation of the photospheric flux from near-IR
to longer wavelengths that determined the detection threshold.
A detection required a disk flux relative to the 
stellar flux at 24 and 70 $\mu$m of $F_{\rm{24disk}}/F_{\rm{24\star}} > R_{\rm{24det}} = 0.1$ 
(for a 5$\sigma$ detection) and $F_{\rm{70disk}}/F_{\rm{70\star}} > R_{\rm{70det}} = 0.55$ (for 
a 3$\sigma$ detection), respectively.
Rearranging equations (\ref{eq:fnustar}) and (\ref{eq:fnudisk2}) shows that such
detection limits correspond to disks with a fractional luminosity of:
\begin{equation}
  f_{\rm{det}} = 6.0 \times 10^9 R_{\rm{det}} r^{-2}L_\star T_\star^{-4} B_\nu(\lambda,T_\star)/
    B_\nu(\lambda,278.3L_\star^{0.25}r^{-0.5}).
  \label{eq:fdet}
\end{equation}

\subsubsection{Fractional luminosity vs radius}
\label{ss:fvsr}

\begin{figure*}
  \centering
  \begin{tabular}{cc}
     \hspace{-0.35in} \includegraphics[width=3.2in]{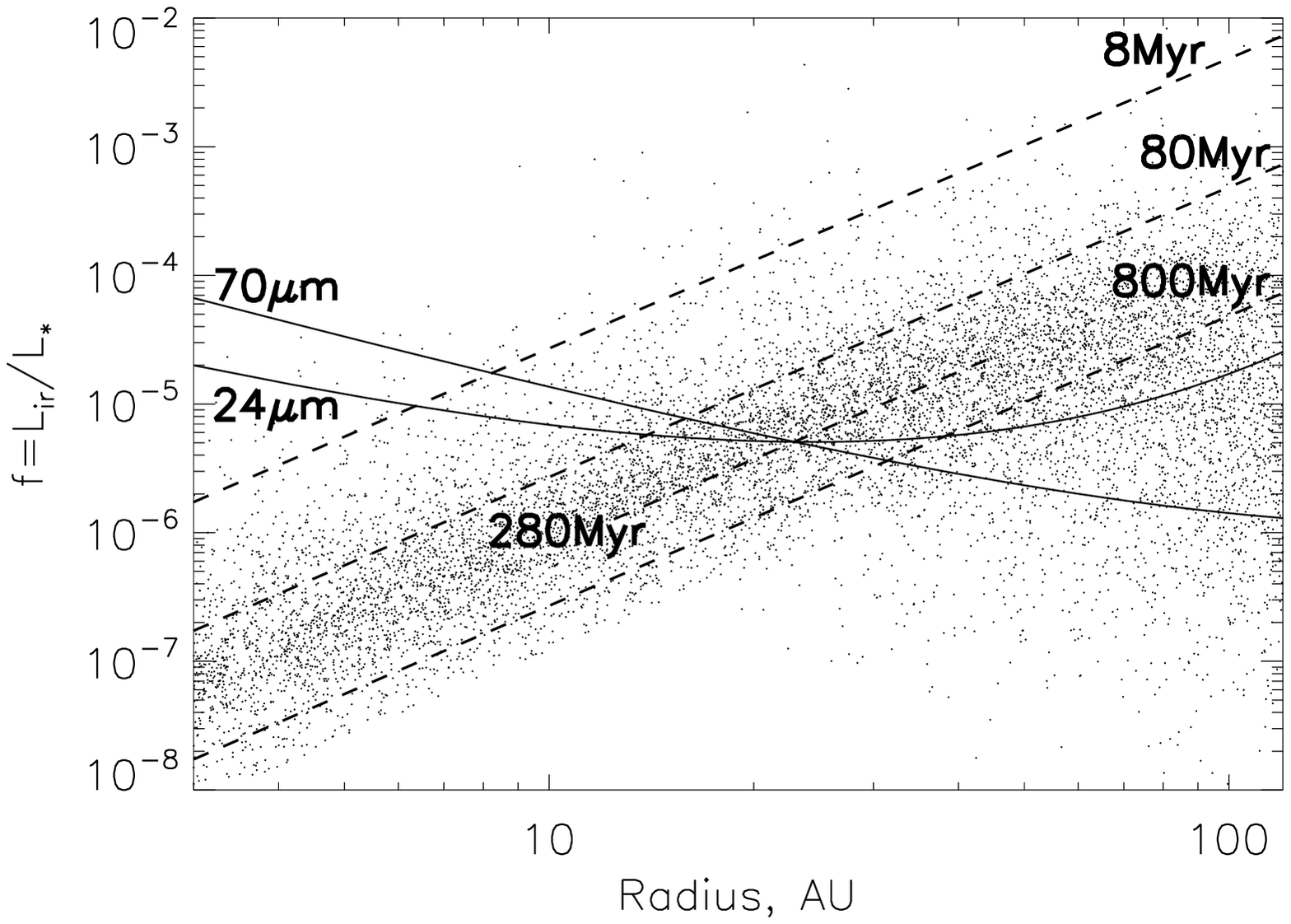} &
     \hspace{-0.35in} \includegraphics[width=3.2in]{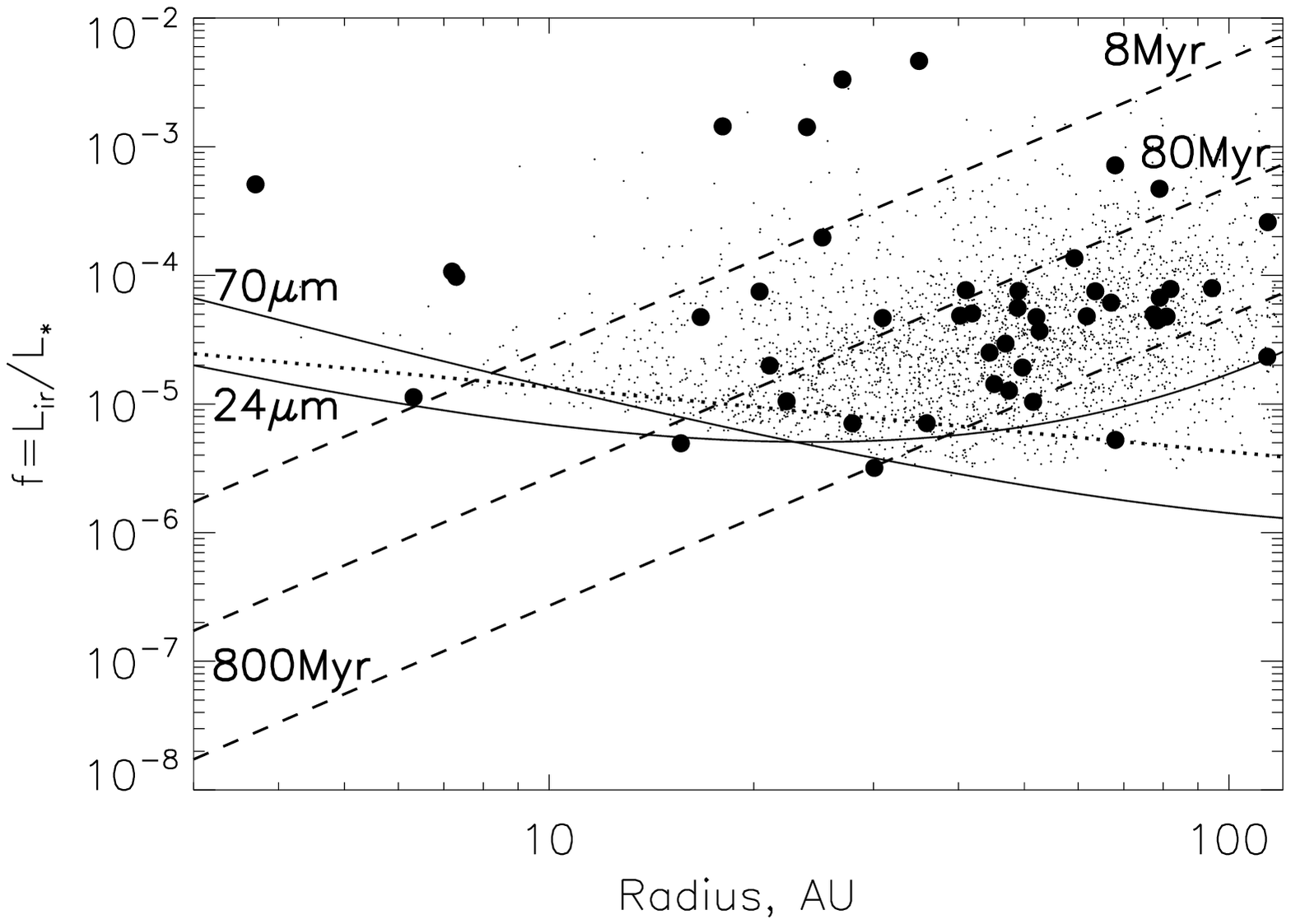}
  \end{tabular}
  \caption{Fractional luminosity versus radius.
  \textbf{(Left)} The 10,000 disks in the A star population model of Fig.~\ref{fig:70um}
  are shown with dots.
  The detection thresholds of the Su et al. (2006) Spitzer observations at 24 and 70 $\mu$m
  ($F_{\rm{24disk}}/F_{\rm{24\star}} > 0.1$ and $F_{\rm{70disk}}/F_{\rm{70\star}} > 0.55$)
  are shown with solid lines for A0V stars (equation \ref{eq:fdet}).
  The maximum luminosities for planetesimal belts around A0V stars of age
  8, 80, 280, and 800 Myr in the model are shown with dashed lines (equation \ref{eq:fmax2}).
  \textbf{(Right)} As (left), except that only disks in the model population that could
  be detected at 24 and 70 $\mu$m are plotted.
  Also shown with filled circles are the properties of the 46 disks that were detected at 
  both 24 and 70 $\mu$m (\S \ref{ss:2470sample}).
  The dotted line shows the limit below which P-R drag becomes important for A0V stars
  (equation \ref{eq:fpr}).
  }
  \label{fig:fvsr}
\end{figure*}

The most valuable plot for interpreting the statistics of debris disk surveys that
have detection thresholds given by equation (\ref{eq:fdet}) is the plot of fractional 
luminosity against radius, Fig.~\ref{fig:fvsr}.
The 10,000 disks in the model population are shown in left Fig.~\ref{fig:fvsr}.
There is a general trend evident in the upper envelope of the model population
following $f \propto r^{7/3}$.
This trend can be understood from equation (\ref{eq:fmax1}), and considering that when 
$X_{\rm{c}} \ll 1$ then $G(11/6,X_{\rm{c}}) \propto r^{-5/6}$, and so
$f_{\rm{max}} \propto r^{7/3}$ (Wyatt et al. 2007).
Lines showing the maximum possible luminosity for planetesimal belts in the model that have ages 
of 8, 80 and 800 Myr, assuming these are around A0V stars with the parameters for 
$Q_{\rm{D}}^\star$, $e$ and $D_{\rm{c}}$ described in \S \ref{s:rieke}, are also
shown on this figure.
These lines are approximately:
\begin{equation}
  f_{\rm{max}} \approx 1.2 \times 10^{-6}r^{7/3}t_{\rm{age}}^{-1}.
  \label{eq:fmax2}
\end{equation}
In other words, the maximum possible luminosity for planetesimal belts of a given radius depends 
only on age, and the upper envelope in the model population is thus determined by the youngest 
age of its disks.

The lines of maximum luminosity are extremely useful, since the evolution on 
Fig.~\ref{fig:fvsr}
of any one planetesimal belt in the model, with initial parameters of radius $r_{1}$ and fractional luminosity 
$f_{\rm{1}}$, is for it to remain stationary at ($r_1$, $f_{\rm{1}}$) while its age is
$t_{\rm{age}} < 1.2 \times 10^{-6}r^{7/3}f_1^{-1}$ Myr, and then for its fractional luminosity to
decrease at later times such that it is always found at ($r_1$, $f_{\rm{max}}$) at an age
$t_{\rm{age}} > 1.2 \times 10^{-6}r^{7/3}f_1^{-1}$ Myr.
This means that a disk of a given age either lies on the appropriate line of maximum luminosity,
or below it, so that, e.g., the sub-sample of model disks that are older than 80 Myr are
(some fraction of) the disks that lie below the line of maximum luminosity for 80 Myr.
Note that while this is useful for illustrating the trends in the data it should not be
applied too rigidly, since these lines are also a function of spectral type, and
their exact location depends on parameters like $Q_{\rm{D}}^\star$, $e$ and 
$D_{\rm{c}}$, which may vary among disks.

Also shown on left Fig.~\ref{fig:fvsr} is the detection limit at 24 and 70 $\mu$m 
appropriate for planetesimal belts around A0V stars assuming
the detection thresholds given in \S \ref{ss:2470sample} (equation \ref{eq:fdet}).
The detection threshold at 24 $\mu$m is roughly flat at $f_{\rm{24det}}=0.5-2 \times 10^{-5}$
across all radii, while that at 70 $\mu$m falls more than an order of magnitude
across the range 3-120 AU.
Disks in the model population that we predict could be detected at 24 and 70 $\mu$m
would lie above both of these lines (noting, however, that the location of the line is
also spectral type dependent).
Such a sub-sample is shown in right Fig.~\ref{fig:fvsr} where also plotted
is the sample of observed disks that were actually detected at 24 
and 70 $\mu$m (\S \ref{ss:2470sample}).
To a large degree the observed population is well described by the model population, both in 
terms of the lower envelope (described by the 24 and 70 $\mu$m detection limits), and the
upper envelope (described by a $\propto r^{7/3}$ power law and close to the line of
maximum luminosity for disks a few Myr old), as well as the distribution of disks
in between, which is more dense between the lines of maximum luminosity for 80 and 800 Myr.
A fit to the population of observed disks gives $f \propto r^{-0.1 \pm 0.3}$ with a large
spread at each radius, which is not inconsistent with that expected from the model population
for disks detected at 24 and 70 $\mu$m ($f \propto r^{0.6}$).
The agreement is closer (fit to observed population of $f \propto r^{0.9 \pm 0.3}$)
when the systems mentioned in the following paragraph are removed.

Closer inspection shows that the disks that lie above the 8 Myr maximum
luminosity line for disks around A0V stars are:
the disks with well known ages in the range 5-12 Myr (HD39060, HD109573, HD110058, HD141569
and HD172555), and three stars that have $f/f_{\rm{max}}$ in the range 30-100 and so
appear unusually bright for their age (HD3003 at 50 Myr, HD38678 at 230 Myr, and HD115892 at
350 Myr; Table \ref{tab:2470sample}).
The presence of the disks of 5-12 Myr stars above the 8 Myr line is to be expected, since
the planetesimal belts in these systems either must be, or are likely to be $<8$ Myr-old.
Even if the star itself is older than 8 Myr, we still need to allow some time for the
planetesimal belt to form, and if this occurs on the same timescale that it takes for
proto-planetary disks to dissipate, then it takes $\sim 6$ Myr (Haisch, Lada, \& Lada 
2001).
Also, at such a young age, there is the possibility that these systems have yet to reach
equilibrium, and that this early phase is characterized by an unusually high quantity of
small dust grains.
Whether the three unusually bright stars are in a transient phase (Wyatt et al. 2007), or have
unusual planetesimal belt properties remains to be seen.

Understanding how the detection limits (equation \ref{eq:fdet}) and constraints from
the maximum possible luminosity for a given age (equation \ref{eq:fmax2}) appear
on the plot of fractional luminosity against radius (left Fig.~\ref{fig:fvsr})
readily explains trends (i) to (iii) given in \S \ref{ss:2470} (Su et al. 2006):
(i) The model predicts that disks that are detected at 70 $\mu$m but not 24 $\mu$m should
lie in the region below the 24 $\mu$m threshold and above the 70 $\mu$m threshold on this plot,
with $f$ in the range $10^{-6}$ to $3 \times 10^{-5}$ (for A0V stars), depending on their radius.
In contrast, those detected at both wavelengths, which should lie above both
thresholds, would have $f$ in the range $5 \times 10^{-6}$ to $10^{-2}$ (for A0V stars). 
Thus the model predicts that the range in $f$ should be systematically higher for disks
detected at both wavelengths than for those detected only at 70 $\mu$m, as found by
Su et al. (2006).
(ii) The region of parameter space occupied by disks that are detected at 70 $\mu$m but not
at 24 $\mu$m also lies entirely below the line of maximum luminosity for 280 Myr (for A0V stars;
Fig.~\ref{fig:fvsr}). 
This means that the only disks in this region of parameter space that are younger than
280 Myr must be those of initially low mass for which the most massive objects within
them have yet to come to collisional equilibrium.
This means that this population has a higher average age (460 Myr) than the rest of the
population (400 Myr).
This is consistent with the findings of Su et al., since the mean age of their group
II sources (i.e., those detected at 70 but not 24 $\mu$m) is higher than average
at $\sim 490$ Myr.
The model can also be used to analyze the population of disks
that are expected to be detected at 24 but not 70 $\mu$m (i.e., Su et al. 2006 
group
V sources).
In the model such disks have small radii ($<23$ AU for A0V stars) and lie above the 280 Myr
line of maximum luminosity (for A0V stars), meaning that there can be no disks in this
population that are older than this limit, and that the mean age of this population is
$\sim 150$ Myr.
Indeed, the oldest star in Su et al. (2006) group V is 400 Myr old, and the mean 
age of
their group V sample is 100 Myr.
\footnote{Some of the Su et al. group V sources were not observed down to the limit 
$R_{\rm{70det}}=0.55$, e.g., due to cirrus confusion.
Thus more sensitive observations may find 70 $\mu$m excesses at $R_{\rm{70det}}>0.55$
in this sample.
However, this would not affect the conclusion that bona-fide group V sources
are, on average, younger than the rest of the population.}
(iii) The lowest excess in the model population is $f \approx 10^{-9}$, and 28\% of
stars in this population have excesses lower than $10^{-7}$, in agreement with non-detections
at this level in Su et al. (2006).

\subsubsection{Fractional luminosity vs age}
\label{ss:fvst}

\begin{figure*}
  \centering
  \begin{tabular}{cc}
     \hspace{-0.35in} \includegraphics[width=3.2in]{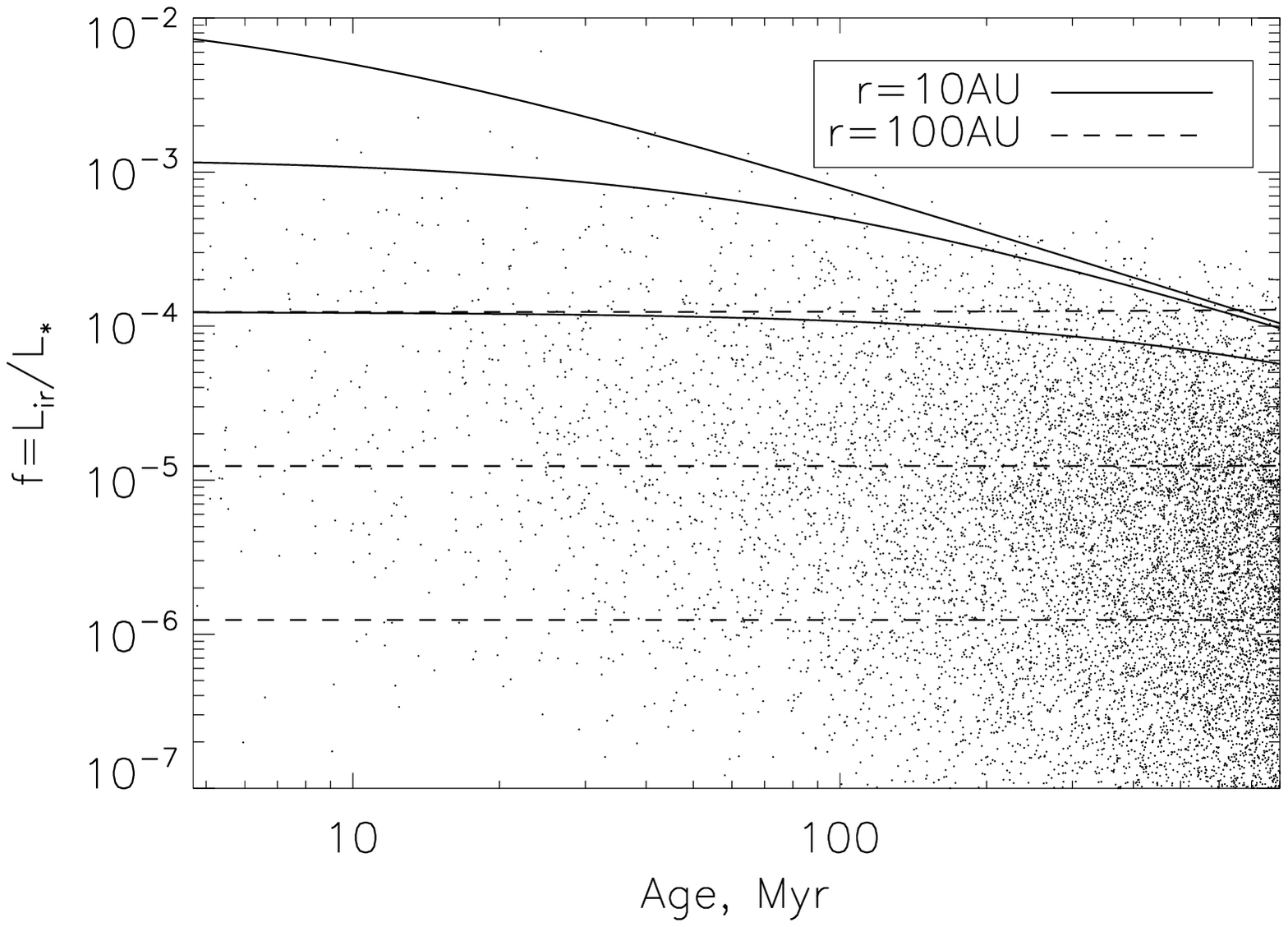} & 
     \hspace{-0.35in} \includegraphics[width=3.2in]{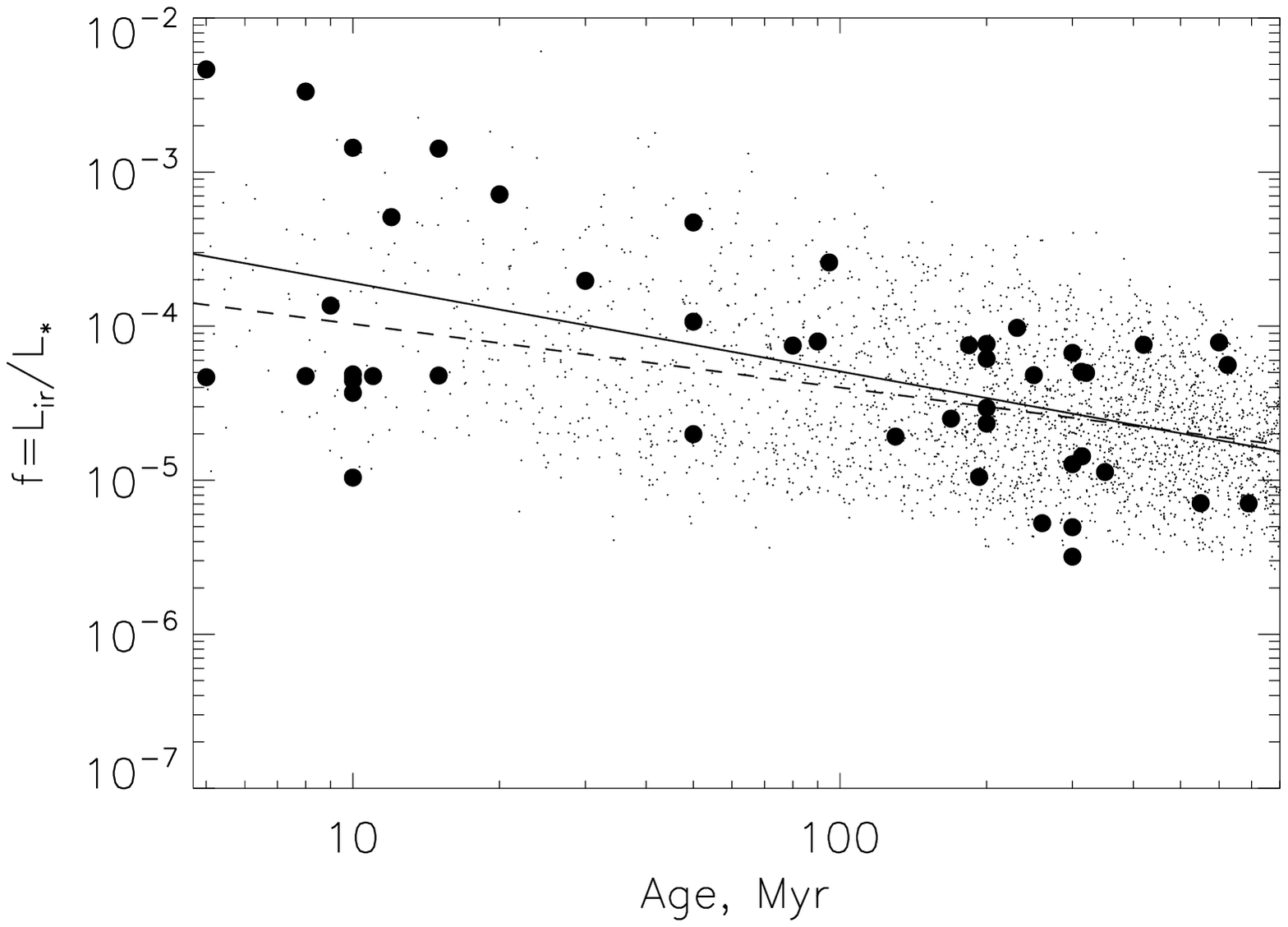}
  \end{tabular}
  \caption{Fractional luminosity versus age.
  \textbf{(Left)} The 10,000 disks in the A star population model of Fig.~\ref{fig:70um} are shown with
  dots.
  The evolutionary tracks of individual planetesimal belts at 10 AU and 100 AU around A0V stars
  are shown with solid and dashed lines, respectively, for starting disk masses of 1, 10 and $100M_\oplus$
  (with higher masses corresponding to higher fractional luminosities at young ages).
  \textbf{(Right)} As (left), except that only disks in the model 
  population that could be detected at 24 and 70 $\mu$m are plotted.
  Also shown with filled circles are the properties of the 46 disks that were detected at
  both 24 and 70 $\mu$m (\S \ref{ss:2470sample}).
  The solid and dashed lines show fits to the observed and model disk populations, respectively.
  }
  \label{fig:fvst}
\end{figure*}

The fractional luminosity of the 10,000 disks in the model population are shown
in left Fig.~\ref{fig:fvst} as a function of stellar age.
Also shown on this plot are lines showing the evolution that a planetesimal
belt at 10 and 100 AU would take assuming starting masses of 1, 10 and $100M_\oplus$.
As described in \S \ref{ss:stst}, this evolution remains flat until the largest planetesimals
collide at which point the fractional luminosity falls off $\propto t_{\rm{age}}^{-1}$.
For the planetesimal belt at 100 AU, the largest planetesimals do not have a chance to
reach collisional equilibrium over the assumed 800 Myr main sequence lifetime of the star,
while for the planetesimal belt at 10 AU, this equilibrium is reached in the range
a few Myr to a few hundred Myr, depending on starting mass.
The upper envelope in the model population can be described as being reasonably flat
at $f \approx 10^{-3}$ until an age of $\sim 100$ Myr at which point it falls off.
The fall-off rate is slightly flatter than $\propto t_{\rm{age}}^{-1}$, which is
because the brightest disks at any given age are at different radii for different
ages.
For example, the upper envelope for ages $<100$ Myr is reasonably well characterized
by the evolution of the $100M_\oplus$ planetesimal belt at 10 AU shown on left
Fig.~\ref{fig:fvst}, whereas the evolution of the $100M_\oplus$ planetesimal
belt at 100 AU has a higher luminosity than that at 10 AU at ages $>600$ Myr. 

The sub-sample of the model population that can be detected at 24 and 70 $\mu$m
is shown in right Fig.~\ref{fig:fvst} and exhibits the same upper
envelope.
The lower envelope of this population also falls off slowly with time, not because
the detection limit drops, but because a smaller fraction of the stars at younger
ages have luminosities approaching that limit, whereas collisional
processing has reduced the fractional luminosities of older stars so that many are
pushed below the detection threshold (particularly for disks at small radii).
The model thus explains trends (iv) and (v) of \S \ref{ss:2470} (Su et al. 
2006):
(iv) The spread in fractional luminosity at any given age is around two orders of
magnitude both in the observed and model populations.
(v) Both the model and observed populations have lower fractional luminosities for
disks around older stars than around younger ones, with the model population exhibiting
a $f \propto t_{\rm{age}}^{-0.39}$ fall-off that is in excellent agreement with the
observed population for which $f \propto t_{\rm{age}}^{-0.57 \pm 0.13}$ (compared with
$f \propto t_{\rm{age}}^{-0.6}$ quoted in Su et al. 2006 for their 
observations).

The biggest difference between the model and observed populations is that the
model assumes that stellar ages are uniformly distributed between 0 and 800 Myr,
which means that the model predicts that 30\% of the population detected at
24 and 70 $\mu$m should be older than 400 Myr, whereas just 11\% of the observed
population are that old.
The most likely cause for this discrepancy is the fact that the mean age of the
stars searched for 24 and 70 $\mu$m excesses is younger than the 400 Myr mean
age of the model population.
The mean age of the Su et al. (2006) sample is 270 Myr which may be attributable
to a bias in the way the sample was chosen, or to the fact that A0V stars reach the
end of the main sequence at $\sim 400$ Myr which means that
a dearth of $>400$ Myr early type stars is to be expected in any sample of main sequence
stars (Greaves \& Wyatt 2003).
The fact that the model provides a close fit to the fraction of stars from any given age
range that lie in specific ranges of fractional excess flux
(\S \ref{ss:bestfitmodel} and \S \ref{ss:70um}) means that the age distribution of 
the observed sample is immaterial suggesting that
a sudden disappearance of disks at $\sim 400$ Myr (Habing et al. 1999)
is not required to explain the statistics.
However, given that 39 stars in Su et al. (2006) were observed in the $>400$ Myr 
population,
and that the model predicts that 46\% (i.e., 18) of these should have disks above the
nominal 70 $\mu$m detection threshold quoted in that paper,
then the observed number of detected disks of 3 is only consistent with the model when one
takes into account that not all of the 39 stars were
observed down to $R_{\rm{70det}} = 0.55$, e.g., because of cirrus confusion.
Almost all (35) of the 39 were observed to $R_{\rm{70det}} = 5$, and the model predicts
that 17\% (i.e., 6) of these should have been detected at this level, which is consistent
with the observed number.
In other words we predict that, unless there is some mechanism destroying these disks at
$\sim 400$ Myr, then more disks remain to be detected around the $>400$ Myr star sample
at the nominal 70 $\mu$m detection threshold.

The evolution of the upper envelope discussed in trend (vi) 
(\S \ref{ss:2470}; Su et al. 2006) is also in agreement with the model, since
we find a fall-off in the upper envelope that is close to $\propto 
t_{\rm{age}}^{-0.9}$.
In answer to the discrepancy with the Decin et al. (2003) result 
that showed a maximum fractional luminosity that is constant with age, 
we first note that this applied to disks around stars of all spectral
types.
When considering their results only for A stars, we find an evolution which is in 
agreement with that of our observed sample, with the exception of HD22128, an A5 star
for which Decin et al. (2003) quote an age of $\sim 1.4$ Gyr and a fractional luminosity
of $70 \times 10^{-5}$ (i.e., which would lie off to the right of our right 
Fig.~\ref{fig:fvst} somewhat above the predicted model population).
We thus consider that the anomalous system HD22128 could either have disk
properties that make it unusually bright, it could be in a transient state, or its
luminosity or age could have been miscalculated.
Indeed an age of 320 Myr has also been quoted for HD22128 (Decin et al. 2003).
Further, recent MIPS observations show that the source is extended at 24 and
70 $\mu$m on $>1000$ AU scales.
Since such emission is unlikely to originate in a disk, this suggests that
the luminosity of the debris disk based on bolometric far-IR observations is
an upper limit and the existence of a debris disk in this system is suspect.

\subsubsection{Radius vs age}
\label{ss:rvst}

\begin{figure}
  \centering
  \begin{tabular}{c}
     \hspace{-0.35in} \includegraphics[width=3.2in]{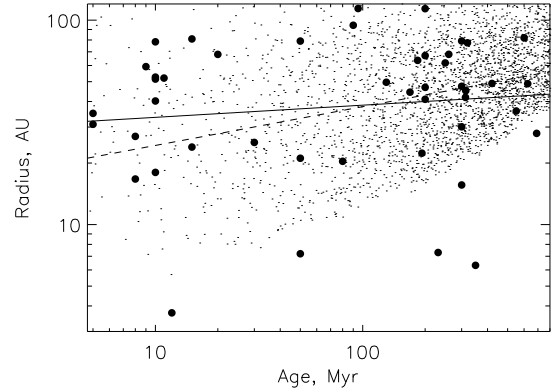}
  \end{tabular}
  \caption{
  Planetesimal belt radius versus age.
  The model population of Fig.~\ref{fig:70um}, shown with dots, includes just those disks that
  could be detected at both 24 $\mu$m or 70 $\mu$m.
  Also shown with filled circles are the properties of the 46 disks that were detected at
  both 24 and 70 $\mu$m (\S \ref{ss:2470sample}).
  Least squares fits to this evolution are shown with solid and dashed lines for the
  observed and model populations, respectively.
  }
  \label{fig:rvst}
\end{figure}

The radii of the 10,000 disks in the model population are shown
in Fig.~\ref{fig:rvst} as a function of stellar age.
Since in this model the radius does not change with age we have plotted
only those disks in the model that it is possible to detect at
24 and 70 $\mu$m and compared these with the properties of the observed
population.
The most noticeable trend of the model population is that
at any given age it is only disks that have radii larger than a certain
limit that can be detected, and that this limit increases with age.
This can be readily understood from left Fig.~\ref{fig:fvsr}.
For example, disks that are both $\sim 80$ Myr old and detected at 24 and 70 $\mu$m
must lie below the 80 Myr line of maximum luminosity on that figure, and
above the 24 and 70 $\mu$m thresholds.
This is a region of parameter space that permits only disks with radii $>16$ AU
(for A0V stars), and that does not extend down so far in radius at later ages
(and encompasses a much smaller area).
Physically this is because planetesimal belts at smaller radii process
their mass faster and so fall below the detection threshold at younger
ages (i.e., it does not mean that planetesimal belts at this radius do
not exist).
This is manifested as an apparent increase in the radius of planetesimal
belts with age, with a least squares fit to the population of detected
disks in the model increasing $\propto t_{\rm{age}}^{0.20}$, although
there is still a large spread in radii at any age.
While the effect is small, the increasing size of the empty region with
age means that the radius distribution (and so the color temperature 
distribution) is broader at younger ages, which means that the
model also reproduces trend (vii) (\S \ref{ss:2470}) seen by Su et al. (2006).
The modest apparent increase in size with age is also consistent with the observed
population, for which we derive a reasonably flat evolution of
$\propto t_{\rm{age}}^{0.06 \pm 0.07}$
(although note that a steeper evolution $\propto t_{\rm{age}}^{0.09 \pm 0.05}$ would be inferred
if the anomalous sources discussed below are removed).
A flat distribution of radii with age was also observed in the sub-mm study of
Najita \& Williams (2005) which included disks around stars of all spectral types.
It was also reported by Rhee et al. (2007), although these authors suggested 
that the
upper envelope in radius may increase across the range 100 to 1000 Myr, an 
observation which is not predicted in our model, and only seen with low 
significance in our observed population.

However, there is a distinct anomaly in that six of the observed disks fall
in the region of Fig.~\ref{fig:rvst} in which there are no
model disks:
HD38678, HD115892, and HD172555 lie well within the forbidden region, while
HD2262, HD3003, and HD106591 lie close to the upper edge of this region.
We offer two possible explanations for these systems.
The first is that these are systems which are undergoing a transient period
of high dust luminosity (Wyatt et al. 2007).
This has already been discussed as a possibility for 
HD38678 in \S \ref{ss:fvsr} based on its high luminosity for
its age (and was also suggested in Moerchen et al. 2007).
The other possibility is that the fitting procedure has underestimated the
radius of the planetesimal belt.

In addition to the possibility that a dust temperature that is higher than
black body has caused the dust location to be underestimated, it is also possible
that the dust is located closer to the star than the planetesimal belt because
of the action of Poynting-Robertson (P-R) drag.
P-R drag becomes important when $10^4 f (r/dr) \sqrt{r/M_\star}/\beta < 1$
(Wyatt 2005), where $\beta$ is the ratio of the radiation force to that of
gravity of the star, since this is when the collisional lifetime of the dust is equal
to the time it would take to reach the star by the drag force.
Assuming, as in the model here, that the dust luminosity is dominated by the 
smallest dust in the distribution with $\beta=0.5$, this limit is given by
$f < f_{\rm{pr}}$, where 
\begin{equation}
  f_{\rm{pr}} = 50 \times 10^{-6}(dr/r)\sqrt{M_\star/r},
  \label{eq:fpr}
\end{equation} 
and this is shown on right Fig.~\ref{fig:fvsr} for A0V stars assuming $dr/r=0.5$,
and compared with the observed fractional luminosity for the 46 disks in Table \ref{tab:2470sample}.
The proximity of this P-R drag limit to the detection threshold at 24 and 70 $\mu$m 
illustrates how P-R drag may become important for disks close to the detection threshold
of Spitzer, at least for early type stars.

The details of how the model would be affected by P-R drag below (or approaching) this level are
not clear without further modeling, but the nature of the drag force means that a lower dust
radius (but not planetesimal belt radius) would be expected.
Thus this is one possible explanation for the close proximity of dust from the stars
HD2262, HD106591, and HD115892, all of which have $f/f_{\rm{pr}} < 1$.
It may be possible to test this observationally, since in the extreme situation where P-R drag dominates
the dust distribution, the surface density of the dust disk would be uniform from the planetesimal
belt all the way in to the star leading to an emission spectrum that increases linearly 
with wavelength ($F_\nu \propto \lambda$; Wyatt 2005).
One other source for which P-R drag may be important in shaping the inner edge of
its dust disk is HD19356, for which $f/f_{\rm{pr}}<1$, while a further 11 disks have
$1<f/f_{\rm{pr}}<5$ (Table \ref{tab:2470sample}).

\subsubsection{Accuracy of $f_{\rm{max}}$}
\label{ss:fmax}

\begin{figure}
  \centering
  \begin{tabular}{c}
     \hspace{-0.35in} \includegraphics[width=3.2in]{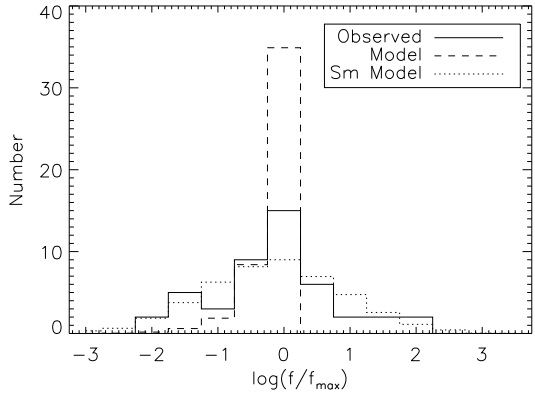}
  \end{tabular}
  \caption{Histogram of the infrared luminosity divided by the maximum
  luminosity possible in the model for a disk of this age, spectral type and radius
  (assuming planetesimal belt properties inferred in \S \ref{s:rsst05pred}).
  The distribution for the observed sample of 46 disks that were detected at
  both 24 and 70 $\mu$m (\S \ref{ss:2470sample}) is shown with a solid line.
  The distribution for disks in the model population of Fig.~\ref{fig:70um} that could be
  detected at 24 and 70 $\mu$m scaled to a population of 46 disks is shown with a dashed line.
  A Gaussian smoothing of 1 dex applied to the model population distribution is shown
  with a dotted line.
  }
  \label{fig:ffmax}
\end{figure}

Figure \ref{fig:ffmax} shows the ratio of observed fractional infrared luminosity
to the maximum luminosity expected for disks at the inferred radius around stars of the age and
spectral type of the hosts; i.e., $f/f_{\rm{max}}$.
This is found for the sample of 46 stars detected at 24 and 70 $\mu$m (\S \ref{ss:2470sample})
to be peaked at $f/f_{\rm{max}} \approx 1$, but extending to around a factor of 100 higher and lower;
the values for this sample are also given in Table \ref{tab:2470sample}.
The sub-sample of the model population detectable at 24 and 70 $\mu$m is also
shown on this figure, and this peaks much more sharply at $f/f_{\rm{max}} \approx 1$, with a
tail to lower $f/f_{\rm{max}}$.
As expected there are no disks in the model with $f>f_{\rm{max}}$, and the fact that
the majority of the model disks have $f \approx f_{\rm{max}}$ indicates that the
majority of the detected disks (in the model) are evolving with their largest planetesimals in
collisional equilibrium.

The observed population has a noticeably broader distribution of $f/f_{\rm{max}}$ than
does the model population.
We interpret this to be a consequence of the assumption that all disks have the same
planetesimal properties.
In reality disks would be expected to have a distribution of properties
and the parameters ($D_{\rm{c}}$, ${Q_{\rm{D}}^\star}$ and $e$) used in the expression
for $f_{\rm{max}}$ would only be representative of the population as a whole rather than
of individual disks.
Thus the breadth of the observed distribution may be indicative of the distribution of
disk properties.
To estimate this breadth we show on the same figure a smoothed model in which the
distribution has been smoothed by a Gaussian of width 1 dex, which fits the observed
distribution well, and suggests a similar breadth in the distribution of
$D_{\rm{c}}^{0.5}{Q_{\rm{D}}^\star}^{5/6}e^{-5/3}$ about the mean values given in
equation (\ref{eq:rel1v2}).
Since the $f/f_{\rm{max}}$ distribution will also have been smeared out due to
uncertainties in the observed (or inferred) disk properties, notably $r$ and
$f$, the true distribution of $D_{\rm{c}}^{0.5}{Q_{\rm{D}}^\star}^{5/6}e^{-5/3}$ is
likely to be much smaller than 1 dex.
It is, of course, also possible that some of the disks inferred to have
$f > f_{\rm{max}}$ are in a transient phase (Wyatt et al. 2007),
which would suggest an even narrower distribution of
$D_{\rm{c}}^{0.5}{Q_{\rm{D}}^\star}^{5/6}e^{-5/3}$ for the
population of disks that are in steady-state.
This indicates a large degree of uniformity in the properties of A star disks.

The most likely candidates for transient emission around A stars are readily
identifed in Table \ref{tab:2470sample} as those with $f/f_{\rm{max}}>10$:
HD3003, HD38678, HD115892, and HD172555.
However, since none of these reaches the threshold $f/f_{\rm{max}} > 100$, we do
not claim that any of these must be transient, but point out that their
properties would have to be extreme compared with the rest of the A star
population for their dust emission to be explained by steady-state collisional
processing of planetesimal belts.\footnote{Note that a size of 3 AU for HD38678
as inferred from mid-IR imaging (Moerchen et al. 2007) would reduce its
maximum possible fractional luminosity given its age, $f_{\rm{max}}$,
by a factor of 7, and so would support the transient nature of this source.}
The inferred breadth of the distribution indicates that a disk with
$f/f_{\rm{max}} > 100$ is a $2\sigma$ deviation, while $f/f_{\rm{max}} > 1000$
is a $3\sigma$ deviation.
We also note that the black body assumption would tend to overestimate the value of 
$f/f_{\rm{max}}$, since underestimating the radius by a factor of 3 would
lead to an overestimation of $f_{\rm{max}}$ by a factor of 13 and so reduce the
factor $f/f_{\rm{max}}$ by an order of magnitude.
In other words, it is not possible to conclude that the disks with
$f/f_{\rm{max}}=10-100$ represent a significant deviation from the disks
found around the remainder of the A star population.

\section{Predictions for SCUBA-2 legacy survey}
\label{s:pred}
While the fact that the model population can explain the available 
far-IR statistics with steady-state collisional evolution is important, this model population 
can also be used to make 
predictions for what we can expect to find in future debris disk surveys.
Since the model was originally constrained by a fit to the 24 $\mu$m statistics, in 
essence the first of those predictions was for the far-IR surveys at wavelengths other 
than 24 $\mu$m, which \S \ref{s:rsst05pred} showed to be confirmed.
This section demonstrates the further applicability of the model, by making
predictions for the outcome of the SCUBA-2 legacy debris disk survey.
This is an unbiased 850 $\mu$m survey of the nearest 100 stars in each of the spectral
types A-M (i.e., 500 stars in total) (Matthews et 
al. 2007), and will be sensitivity limited down to the confusion limit which gives a 
$3\sigma$ limit of $F_{850\rm{lim}} = 0.002$ Jy.
For A stars the survey extends out to around 45 pc.

The model is able to make predictions for the 850 $\mu$m flux of the A star
population through application of equation (\ref{eq:fnudisk2}) to get the
disk flux, as well as using equation (\ref{eq:fnustar}) to get the stellar
flux (although this generally falls below the detection threshold).
Comparing the flux predicted by equation (\ref{eq:fnudisk2}) with that of the
7 A stars in the literature with measured 850 $\mu$m fluxes (HD109573,
Greaves, Mannings \& Holland 2000; HD39060, HD216956, and HD172167,
Holland et al. 1998; HD141569, Sheret et al. 2004;
HD21997 and HD14055, Williams \& Andrews 2006) shows that this equation
overestimates the disk flux by an amount $X_{850}$,
where $X_{850}$ for this population lies in the range 2-8, with a median value
of 4.0.
This is to be expected, since it is known that the emission spectrum falls off steeper
than a black body at wavelengths beyond $\sim 70$ $\mu$m (e.g., Dent et al. 2000),
because of the low emission efficiency of the small dust grains which dominate the
cross-sectional area in the disk.
The factor $X_{850}$ is expected to depend on a combination
of the dust size distribution, compositional properties, planetesimal belt
radius, and stellar spectral type.
Thus, when deriving 850 $\mu$m fluxes for the model population
we reduced the flux found from equation (\ref{eq:fnudisk2}) by a
factor of $X_{850}=4$.
The ability to estimate sub-mm flux with reasonable accuracy from
extrapolation of far-IR data is in agreement with the conclusion of
Rhee et al. (2007) who found a strong correlation of dust masses
derived from sub-mm fluxes, $M_{850}=F_\nu d^2 [\kappa_\nu B_\nu(\lambda,T)]^{-1}$,
with $f$ derived from far-IR fluxes.
Indeed, our model reproduces the trend shown in their Fig. 5,
since with a dust opacity of 0.17 m$^2$ kg$^{-1}$ and a factor of
$X_{850}=4$ our model predicts $f/M_{850} = 14 r^{-2}$ (in units
of $M_\oplus^{-1}$), exactly as observed by Rhee et al. (2007) for
A stars.

\begin{figure}
  \centering
  \begin{tabular}{c}
     \hspace{-0.0in} \includegraphics[width=3.2in]{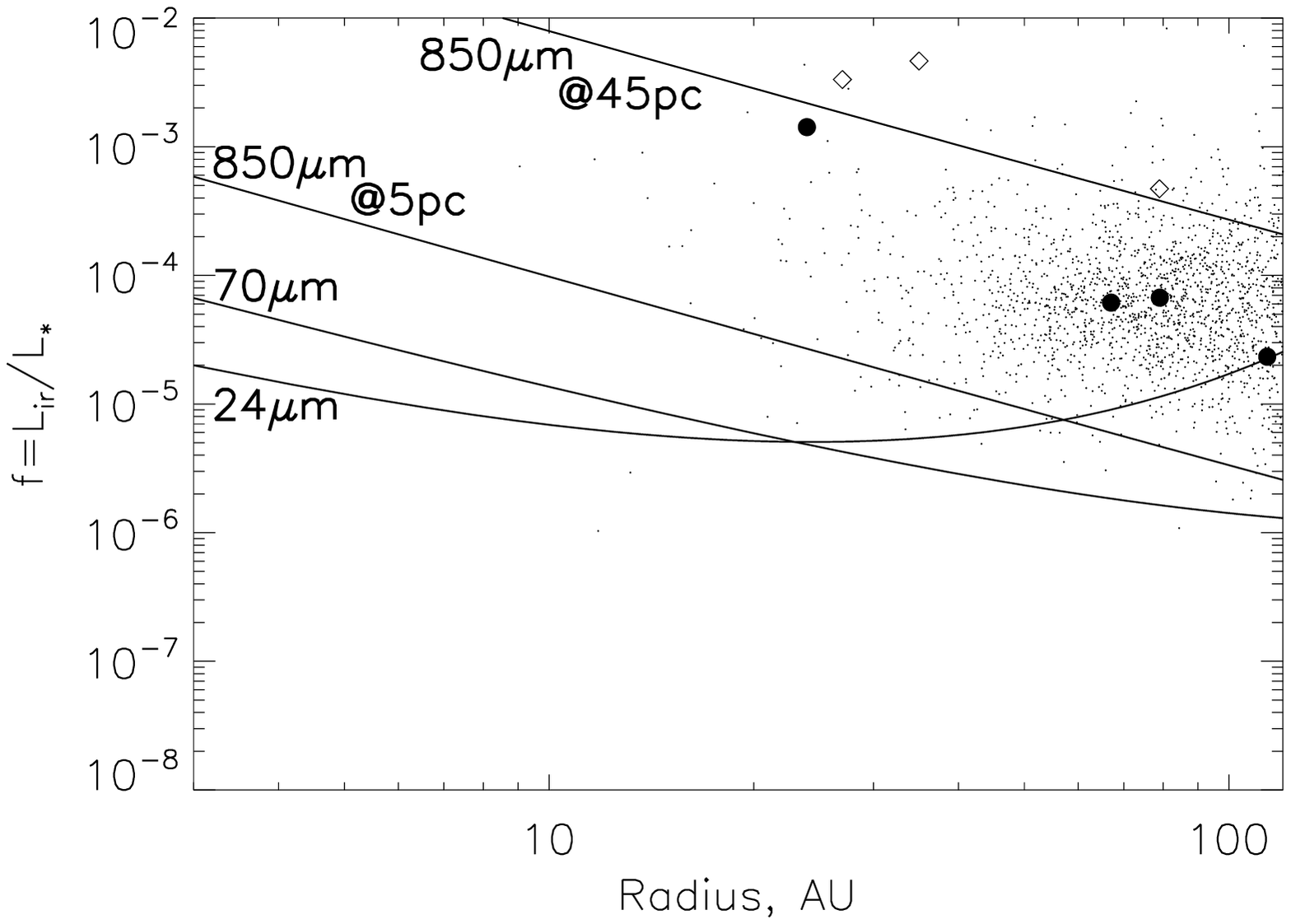} \\[-0.2in]
     \hspace{-0.0in} \includegraphics[width=3.2in]{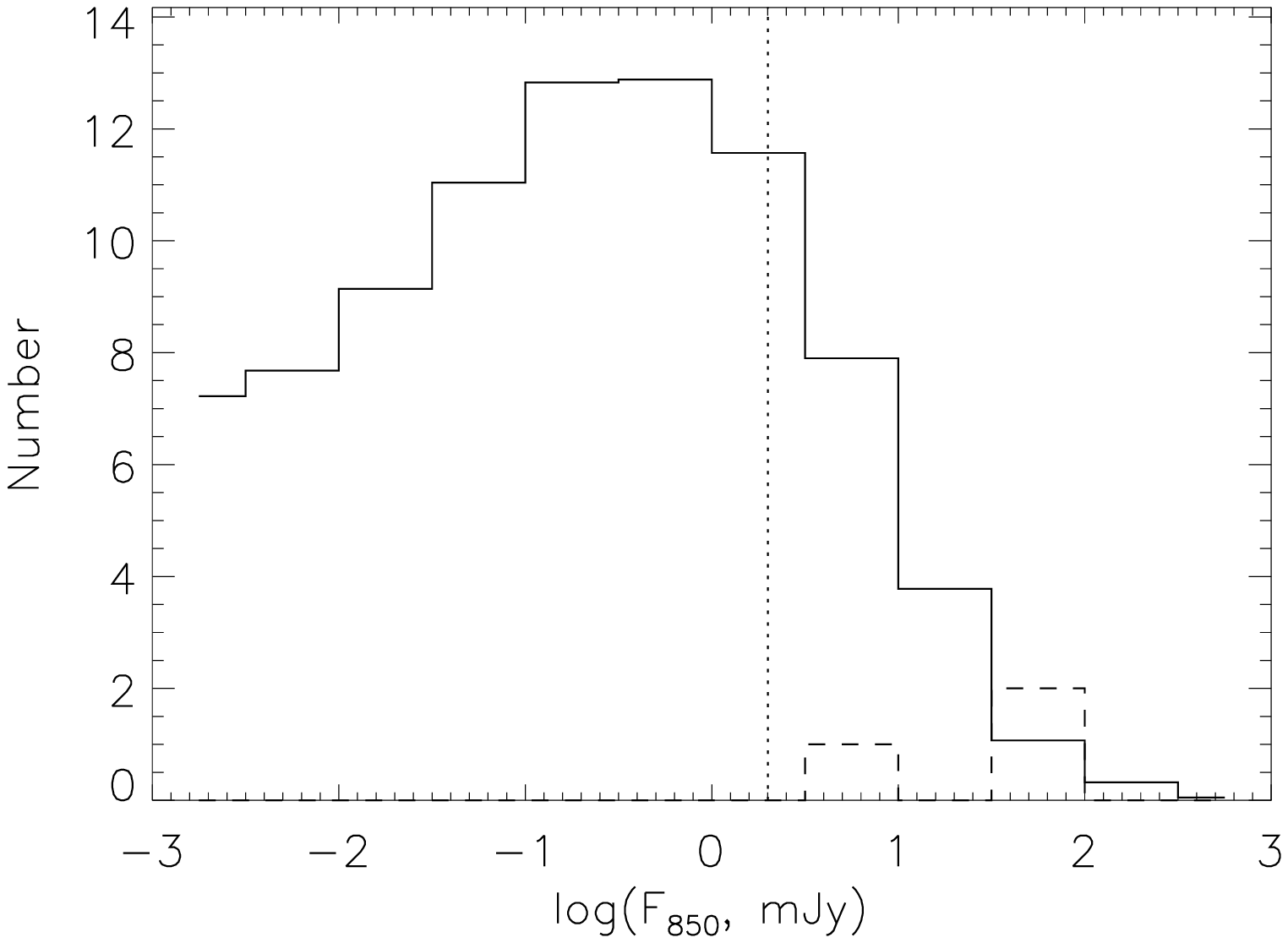}
  \end{tabular}
  \caption{Predictions of the model population of Fig.~\ref{fig:70um}
  for the outcome of the SCUBA-2 legacy survey, an unbiased survey at 850 $\mu$m of the nearest 100 A stars.
  \textbf{(Top)} Fractional luminosity of the sub-sample of the model population which is 
  both within the survey limits ($<42$ pc), and which has a predicted flux above the
  sensitivity limit of the SCUBA-2 survey ($F_{850\rm{disk}}>2$ mJy) (dots).
  The solid lines are the detection thresholds of 24 and 70 $\mu$m Spitzer observations
  of Su et al. (2006), and the 850 $\mu$m detection threshold for disks at 5 and 42 pc
  in the survey;
  all of these detection thresholds apply to disks around A0V stars.
  The disks of A stars at $<42$ pc that have already been detected at 850 $\mu$m are shown with
  filled circles, while those that have been detected but lie beyond 42 pc are shown
  with diamonds.
  \textbf{(Bottom)} Histogram of the distribution of 850 $\mu$m flux expected
  for the 100 A stars within 42 pc.
  The detection limit is shown with a dotted line.
  The dashed line shows the stars in the survey already detected at 850 $\mu$m.
  }
  \label{fig:scuba2}
\end{figure}

Figure \ref{fig:scuba2} shows the predictions of the model for the fractional
luminosity versus radius of the sub-sample of the model population
that have 850 $\mu$m fluxes above the detection threshold of 2 mJy
(noting that all of these are also within the survey threshold of 45 pc).
Also shown are the 7 A star disks with sub-mm detections, indicating the 4
of these that are within 45 pc.
All but one of the 4 within 45 pc will appear in the SCUBA-2 survey;
HD39060 is excluded, since it is below the declination limit of
$-40^\circ$.
There is good agreement between the model and observed populations inside
45 pc, insofar as it is possible to say with such small number statistics.
Since the sub-mm detection limit is determined by the instrumental sensitivity, rather 
than by the accuracy of the photospheric calibration, it is no longer described by
equation (\ref{eq:fdet}), but can be given by
\begin{equation}
  f_{\rm{det}} = 4.3 \times 10^{10} X_{850} F_{850{\rm{lim}}} (d/r)^{2}/
    B_\nu(\lambda,278.3L_\star^{0.25}r^{-0.5}),
  \label{eq:fdet2}
\end{equation}
and so is distance dependent.
This limit falls with planetesimal belt radius, because for disks at a given
distance it is easier to detect those of lower fractional luminosity if they
are at larger distance from the star, since their lower temperature at larger
distances is more than compensated by the larger cross-sectional area of
material.
The SCUBA-2 detection limit is plotted on Fig.~\ref{fig:scuba2} for A0V stars
at 5 and 45 pc assuming their disks have $X_{850}=4$.
The lower envelope of the model population is thus naturally explained by
the small number of A stars that are close to us (of order 14 of the 10,000
are within 5 pc).
It is tempting to suggest that the lower envelope of the detected disks
is explained in the same way, although with just 4 disks the lower
envelope is not so meaningful.
In any case, it is the nearest disks which are detectable in the sub-mm,
while for those in a given distance range it is predominantly large disks,
or those closer in with high fractional luminosity, that can be detected.

Bottom Fig.~\ref{fig:scuba2} shows the histogram of the 850 $\mu$m fluxes of
the model population (noting that this population should be representative
of the SCUBA-2 survey sources, since all are within 45 pc).
The model predicts that 17 of the 100 stars in the SCUBA-2 A star sample
should have a flux above the detection threshold of 2 mJy.
Of those 17: five would be expected above 10 mJy, and 
eight in the range 3-10 mJy, with the remaining four detected with $3-4.5\sigma$
confidence at 2-3 mJy.
The histogram also shows the sub-mm flux distribution for the 3 A stars already
detected, showing that we already know of 2 stars within 45 pc with fluxes
above 10 mJy (HD172167, HD216956), while we know of just one in the
range 3-10 mJy (HD14055).
Thus the SCUBA-2 survey is expected to discover three new bright disks $>10$ mJy,
but can expect to find a further seven new disks at 3-10 mJy levels.
The $>10$ mJy sources are particularly interesting, since at such
a level of flux it is possible to resolve the structure of the
disks (Holland et al. 1998; Wyatt et al. 2005), and
indeed the model predicts that 60\% of these
will have an angular size of $>6$ \arcsec (noting that all may be larger than
this given that the black body assumption underestimates the true disk size by a factor
of up to three, Schneider et al. 2006).
It is also possible to extract some information on the spatial distribution of
the 3-10mJy sources, e.g., giving a direct measure of the size and inclination
of the disk (Greaves et al. 2004), with around 23\% of these predicted to
be larger than 6 \arcsec.

It may even be possible to extract information from the 83 stars without detected
disks, since they are predicted to have a mean flux of 0.60 mJy, of which 0.26 mJy
is the mean disk flux and 0.34 mJy is the mean stellar flux.
By co-adding their data, the statistical noise level would be reduced to
$0.7/\sqrt{83}=0.077$ mJy, suggesting that it should be possible to get a positive
detection of the photospheric emission, and also to discern if there is an excess
coming from the disks with $\sim 3\sigma$ confidence.
However, contamination from extragalactic sources that lie within the SCUBA-2
beam would have a signal of comparable magnitude to the mean disk flux and
would reduce our ability to make a conclusive detection of disk emission from
this population.

The model predicts that there will be little age dependence within the
sample of disks detected by SCUBA-2, in agreement with the result
of Rhee et al. (2007) who estimated the mass evolution based on far-IR
observations.
For example, the model predicts that the mean age of the detected sample is
$\sim 340$ Myr (compared with 400 Myr for the whole population),
with a roughly even number of disks (8-9) detected in both the $<300$ Myr and
$>300$ Myr age bins, which will be indistinguishable from that expected
if there were no age dependence (6 in the $<300$ Myr and 11 in the $>300$ Myr
samples).
Naturally SCUBA-2 is inclined to detect disks with low values
of $X_{850}$, i.e., those for which their emission is more like black body
(which would be inferred to be composed of large grains).
However, given the small range in $X_{850}$ observed, this dependence
is very slight and would affect mainly those disks detected close to the
sensitivity threshold.
There is also predicted to be a small spectral type dependence, in that
19\% of stars later than A3V are predicted to be detected by SCUBA-2, compared
with 15\% of those earlier than A3V.
This arises because, while disks of lower fractional luminosity can be
detected around earlier type stars,
$f_{\rm{det}} \propto L_\star^{-0.25}$ (equation \ref{eq:fdet2}),
the fractional luminosities of their dust belts are also much lower
because of the higher dust blow-out radius,
$f_{\rm{max}} \propto L_\star^{-0.5}$ (equation \ref{eq:fmax1}).
We do not anticipate that this would be detectable within the A star sample,
however, it suggests that the detectability of sun-like stars could be high,
should their disk population be described by the same parameters as for the
A stars.

Top figure \ref{fig:scuba2} also shows the 24 and 70 $\mu$m detection limits of
the Spitzer survey of Su et al. (2006).
This shows that almost without exception (99.9\%) the disks that can be detected
by SCUBA-2 at 850 $\mu$m are also detectable by Spitzer observations at 70 $\mu$m
with a detection threshold of $R_{\rm{70det}}>0.55$, while most (61\%) of those
detectable by SCUBA-2 are also detectable by Spitzer at 24 $\mu$m.
In other words, ten of the 100 A stars in the SCUBA-2 sample should be detectable at
24, 70 and 850 $\mu$m, while seven should be detectable at 70 and 850 $\mu$m,
but not at 24 $\mu$m;
the combination of sub-mm and far-IR data would be very important for
constraining the dust temperature in these disks.
While this suggests that the survey could equally well be done at 70 $\mu$m,
there are three important points to consider.
\textbf{(i)} First, not all of the sample has been observed with Spitzer -
the Su et al. (2006) sample includes 42 sources within 45 pc (of which
20 were detected at 70 $\mu$m), which assuming even coverage across the sky
means that 34 of these should fall in the SCUBA-2 survey (which has declination
limits of $-40^\circ$ and $+80^\circ$).
This implies that two thirds of the SCUBA-2 A star sources have not previously
been searched for disks in the far-IR (although bright disks may
have turned up in IRAS surveys which have a detection threshold
well above that shown in Fig.~\ref{fig:scuba2}).
It is also not possible to scrutinize all stars with Spitzer to the level
of $R_{\rm{70det}}=0.55$ due to cirrus confusion.
Thus it is predicted that the SCUBA-2 survey will turn up a significant number of
genuinely new discoveries.
\textbf{(ii)} Second, it is important to emphasize the unbiased nature of the SCUBA-2
survey.
Several sources have recently been identified as possessing excess emission in the
sub-mm even when no excess is present in far-IR surveys (Wyatt et al. 2003;
Holmes et al. 2003; Najita \& Williams 2005).
These disks must be cool ($<40$K) and at large radius;
e.g., extrapolation of the detection thresholds in Fig.~\ref{fig:scuba2}
to larger radii indicates that the sub-mm is more sensitive than 70 $\mu$m
surveys to disks that are $>210$ AU at 5pc for A0V stars and $>74$ AU at
5 pc for A9V stars.
Since the model population was constrained by 24 and 70 $\mu$m surveys, there
was no need to invoke a population of disks of large radius, although
we know that dust exists around young A stars to hundreds of AU (Kalas \& Jewitt 1995;
Clampin et al. 2003).
Thus the predicted detection statistics should be viewed as a lower limit,
with many large disks being detected at 850 $\mu$m which cannot be
detected in the far-IR.
Previous unbiased sub-mm surveys suggest that $\sim 15$\% of stars could have
disks too cold to detect in the far-IR, which could push the detection
rate to $\sim 30$\% for the SCUBA-2 A star survey.
\textbf{(iii)} Even discounting the ultra-cool disks discussed in (ii), the
sub-mm statistics will set important constraints on the distribution of the
planetesimal belt radii, since the predictions made here are very sensitive
to that distribution.
For example, with a radius distribution $N(r) \propto r^{-0.5}$, which is not
ruled out in the present study at the $1\sigma$ confidence level, the model
would predict that 22 of the 100 A stars would be detected at $>2$ mJy of which 7
would be $>10$ mJy.

\section{Conclusions}
\label{s:conc}
A simple model for the steady-state evolution of dust luminosity for planetesimal
belts evolving due to collisions was described in \S \ref{s:model}.
This section also described how the model could be applied to determine the
properties of the disks of a population of stars, given that these disks would have 
had a range of initial properties.
This was applied in \S \ref{s:rieke} to the population of A stars that was searched
for evidence of dust emission at 24 $\mu$m by Rieke et al. (2005), and it
was shown that their detection statistics as a function of age could be
reproduced with a model population in which the largest planetesimals
have a size of $D_{\rm{c}}=60$ km, and all planetesimals have a strength of 
$Q_{\rm{D}}^\star = 300$ J kg$^{-1}$ and an eccentricity of $e=0.05$.
However, these parameters should not be over-interpreted, since the observable
properties of the model population remain the same as long as the combination
$D_{\rm{c}}^{0.5}{Q_{\rm{D}}^\star}^{5/6}e^{-5/3}$ is unchanged.
Thus more detailed models of the collisional evolution of planetesimal belts
(e.g., Krivov, Sremcevi\'{c} \& Spahn 2005) are needed to interpret these parameters, e.g., in terms of
the composition of the planetesimal belts.
For now we note that these are reasonable parameters based on planetesimal
(Benz \& Asphaug 1999) and planet formation models (Kenyon \& Bromley 2002).
The model used as input a distribution of planetesimal belt starting masses that
is log-normal with the same width ($1\sigma$ of 1.14 dex) as inferred for
proto-planetary disks (Andrews \& Williams 2005), and found such a distribution
would have to be centred on $M_{\rm{mid}}=10M_\oplus$.
This mass is consistent with that expected for proto-planetary disks around A stars, 
although the observable properties of the model population are also reproduced as
long as $M_{\rm{mid}}D_{\rm{c}}^{-0.5}$ is unchanged.
In other words, the observable 24 $\mu$m properties of main sequence A star disks
can be explained by the steady-state evolution of a disk population with realistic
starting parameters (although more extreme solutions are also possible).

In \S \ref{s:rsst05pred} this model was tested against its predictions for
the 70 $\mu$m properties of the disks in this population.
These were found to reproduce the statistics of the Su et al. 
(2006) survey which showed a much longer decay timescale for the
70 $\mu$m excess than for that at 24 $\mu$m.
The model was also used to predict the properties of the sub-sample 
of disks that could be detected at both 24 and 70 $\mu$m, and this was
compared with a sample of 46 A stars compiled from the literature
for which excess emission has been detected at both 24 and 70 $\mu$m.
The model reproduces and explains the distribution of
the observed disks on the $f$ vs $r$ (Fig.~\ref{fig:fvsr}), $f$ vs $t_{\rm{age}}$
(Fig.~\ref{fig:fvst}) and $r$ vs $t_{\rm{age}}$ (Fig.~\ref{fig:rvst})
plots, including an upper envelope in the $f$ vs $r$ plot that increases $\propto 
r^{7/3}$,
a decay in the mean luminosity of disks with age $\propto t_{\rm{age}}^{-0.39}$,
and a mean radius of detected disks that, if anything, increases with
age $\propto t_{\rm{age}}^{0.20}$.

Thus it appears that the wide range in typical disk properties can be explained
without appealing to stochasticity.
The large spread in 24 $\mu$m fractional excess at any age (Rieke et al. 2005)
can occur naturally from a spread in initial disk masses and radii, and
the slower fall-off at 70 $\mu$m and more detailed distributions of
the properties of disks detected at 24 and 70 $\mu$m (Su et al. 2006) could
be the consequence of the detection bias.
This bias is best explained on the plot of $f$ vs $r$, on which it is
possible to plot the lines above which disks must lie to be detected
at 24 and 70 $\mu$m, and below which they must lie for a given
age.
One illustrative example of this bias is the fact that while in the
model the planetesimal belt of any individual star does not change in
radius, the mean radius of detected planetesimal belts is predicted to
increase with age.
The reason is that planetesimal belts that are closest to the star
are processed faster and so fall below the detection threshold much
faster than planetesimal belts that are farther from the star.
Thus if such a trend is seen it would not necessarily be evidence
for delayed stirring, in which the collisional cascade is only
initiated once enough time has elapsed for Pluto-sized objects to
form, a process which takes longer further from the star (e.g.,
Kenyon \& Bromley 2002; Dominik \& Decin 2003).
Rather it could be evidence that the processing of the planetesimal
belt is occurring inside-out, as inferred by Su et al. (2006).

This does not rule out the possibility that some, or even all, A star
debris disks in the Rieke et al. (2005) and Su et al. (2006)
samples are undergoing transient events on top of the overall trend
established by the model.
In fact, there are a few disks that may (rather than must) require
this mechanism to explain their properties.
The potentially transient systems identified in this paper are:
HD3003, HD38678, and HD172555, which have high luminosity for their
age;
as well as HD115892, which has unusually an low radius for its
age.
All of these systems are identifiable by their high value of $f/f_{\rm{max}}$,
where $f_{\rm{max}}$ is the maximum possible fractional luminosity that
a disk can have given its age and radius (assuming the planetesimal
belt properties inferred for the rest of the population).
We also found that the disks of HD2262 and HD106591 have unusually low
radii for their age, and suggested that this may be attributable to the
action of P-R drag on dust produced in their dust belts, since this starts
to become important when $f<f_{\rm{pr}}$, and these belts (along with HD115892
and HD19356) can be identified from the sample as having $f/f_{\rm{pr}}<1$.
Neither do the results in this paper imply that the collisional cascade 
in all A star planetesimal belts must have been initiated soon ($\sim 10$ Myr) 
after the stars formed, rather than the collisional cascade being initiated 
after a delay of several 10s or 100s of Myr in some systems (i.e., the delayed 
stirring model).
Indeed, if the lack of large radii disks at young ages is confirmed (Rhee
et al. 2007), this would support some role for delayed stirring in debris
disk evolution.

Analysis of the distribution of $f/f_{\rm{max}}$ of the 46 A star
disks detected at both 24 and 70 $\mu$m shows a distribution that
may be attributed to the planetesimal belts not having exactly the same
properties, but having a distribution in the combination of parameters given by
$D_{\rm{c}}^{0.5}{Q_{\rm{D}}^\star}^{5/6}e^{-5/3}$ (which is also the
combination in the parameter $f_{\rm{max}}$) that is log-normal
with a $1\sigma$ width of 1 dex.
This means two things:
First, that the distribution of parameters of these A stars is quite uniform.
This is perhaps to be expected if the planetesimals in all belts are
of similar composition, and grow to a similar maximum size before perturbations
from the largest member of the cascade (around 2000 km) stir the rest of the
planetesimal belt resulting in eccentricities of similar magnitude
(e.g., Kenyon \& Bromley 2002).
Second, that the anomalous systems identified in the last paragraph
are only $1-2\sigma$ anomalies in that they may be explained as
disks with higher than average planetesimal strength or size,
or lower than average eccentricity.
In contrast, this analysis strengthens the result of Wyatt et al. (2007),
in which the dust emission from several sun-like stars, such as HD69830,
was inferred to be transient, since if the planetesimal belts of
sun-like stars are described by similar properties to those of A stars,
then all of the disks they inferred to be transient have
$f/f_{\rm{max}} > 4000$ and each one would be a $>3.6\sigma$ anomaly
(and yet such disks are found around $\sim 2$\% of stars).
However, given current limits in definitive detection of transient
events, the derived incidence is probably a lower limit.

One particularly interesting outcome of the model is the distribution
of radii of the planetesimal belts.
The model assumed a power law distribution of radii in the
range 3-120 AU, and found from the distribution of 
radii of the sample detected at 24 and 70 $\mu$m, that
this follows $N(r) \propto r^{-0.8 \pm 0.3}$.
It was found to be important to take the 
detection bias into account when interpreting the observed distribution.
The origin of this distribution is not considered in this model,
which assumes that all stars have relatively narrow planetesimal belts
with $dr/r=0.5$.
The inner hole in these belts has been inferred in other studies to be
caused by the presence of inner planets (e.g., Roques et al. 1994;
Wyatt et al. 1999; Wilner et al. 2002; Wyatt 2003; Quillen 2006).
If these are Kuiper belt analogues (Wyatt et al. 2003), then this radius
distribution would be indicative of the orbital radius of the outermost
planet in its planetary system.
However, these could also be asteroid belt analogues, i.e., planetesimal
belts in the midst of a planetary system.
If so the interpretation of this radius distribution is less clear, and could,
like the solar system for example, be indicative of the orbital radius of the
most massive giant planet in these systems.
In any case, this distribution provides an important and unique constraint
on the outcome of planet formation models.

Another particularly powerful application of the model is that, regardless
of the physical reason, it does explain the far-IR properties of A star disks
and so it can also be used to make predictions for what surveys at other
wavelengths (or with different detection thresholds) will see.
Such predictions are made much easier when the survey detection threshold
is uniform, or at least can be readily characterized, and this study
emphasizes the importance of a characterizable threshold for future surveys.
One such survey is the SCUBA-2 debris disks survey (Matthews et al. 2007),
which will include an unbiased sub-mm survey of the 100 nearest A stars to a
uniform $3\sigma$ sensitivity of 2 mJy at 850 $\mu$m.
The model predicts that 17 of the 100 will be detected above 2 mJy,
including 5 above 10 mJy, and that 5 of the detected disks should be resolvable
in sub-mm imaging on size scales $>6$\arcsec.
However, the detected fraction could be higher than this, since it depends on
the radius distribution, which these observations would set strong constraints
on.
It is also unknown how many A stars possess disks too cold to detect in the
far-IR (Wyatt, Dent \& Greaves 2003; Najita \& Williams 2005).
This suggests that the SCUBA-2 survey will be particularly fruitful, and
the model presented here provides a framework which can be used to interpret
the results of this survey, and those of future surveys.

\acknowledgements
We are grateful for support provided by the Royal Society (MCW),
PPARC (RS, JSG) and SUPA (JSG).
This work was partially supported by JPL/Caltech contract 1255094 to the
University of Arizona.


\end{document}